\documentclass[10pt,a4paper]{article}

\usepackage{times}
\usepackage[centertags]{amsmath}
\usepackage{amssymb,amsfonts,theorem,stmaryrd}
\usepackage{slashbox}
\usepackage{pstricks}

\allowdisplaybreaks

\theorembodyfont{\slshape}
\newtheorem{thm}{Theorem}[section]
\newtheorem{prop}[thm]{Proposition}

\newtheorem{lemma}[thm]{Lemma}
\newtheorem{cor}[thm]{Corollary}
\theorembodyfont{\rmfamily} 
\newtheorem{Def}[thm]{Definition}
\newtheorem{rem}[thm]{Remark}

\newenvironment{pf}{\textit{Proof. }}{\hfill$\boxbox$}

\newcommand{\R}{\mathbb{R}}

\newcommand{\Z}{\mathbb{Z}}
\newcommand{\T}{\mathcal{T}}
\newcommand{\A}{\mathcal{A}}
\newcommand{\HH}{\mathcal{H}}
\newcommand{\OO}{\operatorname{O}}
\newcommand{\Riem}{\operatorname{Riem}}
\newcommand{\riem}{\operatorname{riem}}
\newcommand{\Ric}{\operatorname{Ric}}
\newcommand{\ric}{\operatorname{ric}}
\newcommand{\B}{\operatorname{b}}
\newcommand{\BB}{\operatorname{B}}
\newcommand{\CC}{\operatorname{C}}
\newcommand{\cc}{\operatorname{c}}

\newcommand{\tr}{\operatorname{tr}}

\newcommand{\dvol}{\operatorname{dvol}}
\newcommand{\vol}{\operatorname{vol}}
\newcommand{\grad}{\operatorname{grad}}
\newcommand{\divergenz}{\operatorname{div}}

\newcommand{\bb}{\begin{eqnarray}}
\newcommand{\ee}{\end{eqnarray}}
\newcommand{\eee}{\nonumber\end{eqnarray}}

\begin{document}
\title{On Gravity, Torsion and the Spectral Action Principle}
\author{Frank Pf\"aff\-le  \& Christoph A.Stephan \\[.1cm] Institut f\"ur Mathematik \\ Universit\"at Potsdam \\ Am Neuen Palais 10 \\ 14469 Potsdam, Germany}  

\maketitle

\begin{abstract}
\noindent
We consider compact Riemannian spin manifolds without boundary equipped with orthogonal connections.
We investigate the induced Dirac operators and the associated commutative spectral triples.
In case of dimension four and totally anti-symmetric torsion 
we compute the Chamseddine-Connes spectral action, deduce the equations of motions and discuss  critical points.
\end{abstract}


\section{Introduction}
Einstein's field equations can be deduced as equations of motion of the Einstein-Hilbert functional.
In the classical context one considers (Lorentzian or Riemannian) manifolds equipped only with the Levi-Civita connection.\medskip

\noindent
In the 1920s \'E.~Cartan investigated general orthogonal connections i.e.~connections which are compatible with the metric.
The difference of such a connection and the Levi-Civita connection is called torsion.
In his seminal articles \cite{Cartan23}, \cite{Cartan24} and \cite{Cartan25} Cartan observed that in general the torsion tensor splits into three components: the vectorial torsion, the totally anti-symmetric one and the one of Cartan type.
Taking the scalar curvature of orthogonal connections as the Langrangian one attains the Einstein-Cartan-Hilbert functional.
Its critial points are exactly Einstein manifolds, in particular the torsion of a critical point is zero. 
(Physics' literature refers to this fact as the Palatini formalism.)\medskip

\noindent
Aiming for a unified theory of gravity and the other forces Chamseddine and Connes introduced the spectral action principle (\cite{ConnesChamseddine1}).
It states that any reasonable physical action is determined only by the spectrum of a Dirac operator.
Specifically, the Chamseddine-Connes spectral action comprises the Einstein-Hilbert action and the full bosonic part of the action of the Standard Model of Particle Physics, if one considers  suitable twisted Dirac operators based on Levi-Civita connections.
It even predicts the correct Higgs potential necessary for the electro-weak symmetry breaking and allows to put constraints on the Higgs mass.
Into this framework orthogonal connections with totally anti-symmetric torsion have been incorporated in \cite{Torsion}, and by Iochum, Levy and Vassilevich in \cite{IochumLevy} for the purely gravitational action on manifolds with boundary.
Restricting to connections with totally anti-symmetric torsion was geometrically justified by the fact that the geodesics of such connections coincide with those of the Levi-Civita connection.\medskip

\noindent
Throughout the present article we consider closed manifolds, i.e. compact ones without boundary. For these we deal with the full class of orthogonal connections.
We review Cartan's classification and Einstein-Cartan theory in section~\ref{section1}, and we compute some curvature quantities in the case of totally anti-symmetric torsion in section~\ref{section2}.
In section~\ref{section3} we describe the Dirac operators constructed from orthogonal connections, and we notice that the vectorial component of the torsion has to be zero to assure that the Dirac operator is symmetric. This follows from a result by Friedrich and Sulanke (\cite{FriedrichSulanke}).
We show that the Cartan type component of the torsion has no effect on the Dirac operator (even pointwise) which provides another good reason to consider only anti-symmetric torsion.
\medskip

\noindent
In this setting many examples for commutative geometries in the sense of Connes' spectral triples (\cite{Connes96}) can be supplied.
Given the Reconstruction Theorem (\cite{Connes08}), we remark that in the even-dimensional case anti-symmetric torsion is reconstructable from the spectral data.
In four dimensions we calculate the purely gravitational part of the Chamseddine-Connes spectral action in some detail. 
We find that some terms of the action given in \cite{Torsion} actually vanish, thus confirming the result by \cite{IochumLevy}.
For this action we derive the equations of motion in Theorem~\ref{Equation_of_Motion}.
One of them is a Proca equation for the torsion $3$-form which suggests an interpretation of the torsion as massive vector boson.
The set of critial points of the action, i.e.~the solutions of the equations of motion, contains all Einstein manifolds (with zero torsion).
Furthermore, in Lemmas~\ref{lemma_Robertson_Walker} and~\ref{warpedtorsiontorus} we exclude critical points which are warped products and carry special choices of non-zero torsion.
\medskip

\noindent
We tried to keep this text elementary and accessible, and we hope that it may also serve as an introduction for non-experts.
\medskip

\noindent
{\bf Acknoledgement:} The authors appreciate funding by the Deutsche Forschungsgemeinschaft, in particular by the SFB {\it Raum-Zeit-Materie}.
We would like to thank Christian B\"ar and Thomas Sch\"ucker for their support and helpful discussions.

\section{Orthogonal connections on Riemannian manifolds}\label{section1}
We consider an $n$-dimensional manifold $M$ equipped with some Riemannian metric $g$.
Let $\nabla^{g}$ denote the Levi-Civita connection on the tangent bundle.
For any affine connection $\nabla$ on the tangent bundle there exists a $(2,1)$-tensor field $A$ such that
\begin{equation}\label{def_Zusammenhang_mit_Torsion}
\nabla_X Y= \nabla^g_X Y +A(X,Y)
\end{equation}
for all vector fields $X,Y$.\medskip

\noindent
In this article we will require all connections $\nabla$ to be {\it orthogonal}, i.e.~for all vector fields $X,Y,Z$ one has
\begin{equation}\label{def_metrisch}
\partial_X \left\langle Y,Z \right\rangle=\left\langle \nabla_XY,Z\right\rangle+\left\langle Y,\nabla_XZ\right\rangle,
\end{equation}
where $\left\langle\cdot,\cdot\right\rangle$ denotes the scalar product given by the Riemannian metric $g$.
For any tangent vector $X$ one gets  from (\ref{def_Zusammenhang_mit_Torsion}) and (\ref{def_metrisch}) that the endomorphism $A(X,\cdot)$ is skew-adjoint:
\begin{equation}\label{A_schiefadjungiert}
\left\langle A(X,Y), Z\right\rangle = -\left\langle Y,A(X, Z)\right\rangle.
\end{equation}

\noindent
Next, we want to express some curvature quantities for $\nabla$ in terms of $A$ and curvature quantities for $\nabla^g$.
To that end we fix some point $p\in M$, and we extend any tangent vectors $X,Y,Z,W\in T_p M$ to vector fields again denoted by $X,Y,Z,W$ being {\it synchronous} in $p$, which means
\begin{equation*}
\nabla^g_VX=\nabla^g_VY=\nabla^g_VZ=\nabla^g_VW=0 \quad\mbox{ for any tangent vector }V\in T_pM.
\end{equation*}
Furthermore, we choose a local orthogonal frame of vector fields $E_1,\ldots,E_n$ on a neighbourhood of $p$, all being synchronous in $p$.
Then the Lie bracket $[X,Y]=\nabla^g_XY-\nabla^g_YX=0$ vanishes in $p$, and synchronicity in $p$ implies
\begin{equation*}
\nabla_X\nabla_YZ 
=\nabla^g_X\nabla^g_YZ 
+\left(\nabla^g_X A \right)(Y,Z) + A\left(X,A(Y,Z)\right)
\end{equation*}
Hence, in $p$ the Riemann tensor of $\nabla$ reads as 
\begin{align}
\label{Riemann_Torsion_Allgemein}
\Riem(X,Y)Z&=\nabla_X\nabla_YZ-\nabla_Y\nabla_XZ -\nabla_{[X,Y]}Z\nonumber \\
&=\Riem^g(X,Y)Z +\left(\nabla^g_X A \right)(Y,Z) -\left(\nabla^g_Y A \right)(X,Z)
+ A\left(X,A(Y,Z)\right)-A\left(Y,A(X,Z)\right)
\end{align}
where $\Riem^g$ denotes the Riemann tensor of $\nabla^g$.
We note that $\Riem(X,Y)Z$ is anti-symmetric in $X$ and $Y$.
And by differentiation of (\ref{A_schiefadjungiert}) we get that $(\nabla^g_{E_i} A )(E_j,\cdot)$ and $(\nabla^g_{E_j} A )(E_i,\cdot)$ are skew-adjoint, and therefore we have
\begin{equation}\label{Riemann_Torsion_antisymm_in_34_indices}
\left\langle\Riem(E_i,E_j)E_k,E_l\right\rangle = -
\left\langle\Riem(E_i,E_j)E_l,E_k\right\rangle.
\end{equation}
In general, $\Riem$ does not satisfy the Bianchi identity.
The Ricci curvature of $\nabla$ is defined as
\begin{equation*}
\ric(X,Y)=\tr\,\left( V\mapsto \Riem(V,X)Y\right),
\end{equation*}
by (\ref{Riemann_Torsion_Allgemein}) this can be expressed as
\begin{eqnarray}
\ric(X,Y)&=& \sum_{i=1}^n \left\langle \Riem(E_i,X)Y,E_i\right\rangle \nonumber\\
&=& \ric^g(X,Y) 
+\sum_{i=1}^n \left(
\left\langle
\left(\nabla^g_{E_i}A \right)(X,Y),E_i
\right\rangle
-
\left\langle
\left(\nabla^g_{X}A \right)(E_i,Y),E_i
\right\rangle
\right)\nonumber\\
&&\qquad\qquad
+\sum_{i=1}^n\left(
-\left\langle 
A(X,Y),A(E_i,E_i)
\right\rangle
+\left\langle 
A(E_i,Y),A(X,E_i)
\right\rangle
\right)
\label{Ricci_Torsion_Allgemein}
\end{eqnarray}
where $\ric^g$ is the Ricci curvature of $\nabla^g$.
We have used that $A(E_i,\cdot)$ and $A(X,\cdot)$ are skew-adjoint.
\medskip

\noindent
One obtains the scalar curvature $R$ of $\nabla$ by taking yet another trace, 
in $p$ it is given as $R=\sum_{j=1}^n \ric(E_j,E_j)$.
For the following calculation we use that $(\nabla^g_V A)(X,\cdot)$ is skew-adjoint for any tangent vectors $V,X$, 
and we get:
\begin{eqnarray}
R&=&R^g+
\sum_{i,j=1}^n
\left( 
\left\langle 
\left(\nabla^g_{E_i}A \right)(E_j,E_j),E_i
\right\rangle
+
\left\langle 
E_j, \left(\nabla^g_{E_j}A \right)(E_i,E_i)
\right\rangle
\right)
\nonumber\\
&& \qquad
+
\sum_{i,j=1}^n
\left( 
-\left\langle A(E_j,E_j),A(E_i,E_i)
\right\rangle
+
\left\langle A(E_i,E_j),A(E_j,E_i)
\right\rangle
\right)\nonumber\\
&=& R^g+ 2\, \sum_{i,j=1}^n
\left\langle 
\left(\nabla^g_{E_i}A \right)(E_j,E_j),E_i
\right\rangle
-
\big\|\sum_{i=1}^n A(E_i,E_i) \big\|^2
+
\sum_{i,j=1}^n
\left\langle A(E_i,E_j),A(E_j,E_i)
\right\rangle
\label{Scalar_Allgemein}
\end{eqnarray}
where $R^g$ denotes the scalar curvature of $\nabla^g$.\medskip

\noindent
The classification of orthogonal connections with torsions traces back to \cite[Chap.~VIII]{Cartan25}.
Here we adopt the notations of \cite[Chap.~3]{Tricerri} (see also \cite{AgricolaSrni}).
From (\ref{A_schiefadjungiert}) we know that the torsion tensor $A(X,\cdot)$ is skew-adjoint on the tangent space $T_pM$.
Any torsion tensor $A$ induces a $(3,0)$-tensor by setting
\begin{equation*}
A_{XYZ} =\left\langle  A(X,Y),Z \right\rangle\qquad\mbox{ for any } X,Y,Z\in T_pM.
\end{equation*}
We define the space of all possible torsion tensors on $T_pM$ by
\begin{equation*}
\T(T_pM)=\left\{ A\in {\bigotimes}^3T^*_pM\; \big| \; A_{XYZ}=-A_{XZY}\quad\forall X,Y,Z\in T_pM \right\}.
\end{equation*}
This vector space carries a  scalar product
\bb
\left\langle A,A' \right\rangle= \sum_{i,j,k=1}^n A_{E_i E_j E_k} A'_{E_i E_j E_k},
\label{30tensorprod} 
\ee
and the orthogonal group $O(T_pM)$ acts on $\T(T_pM)$ via $(\alpha A)_{XYZ}=A_{\alpha^{-1}(X)\alpha^{-1}(Y)\alpha^{-1}(Z)}$.\medskip

\noindent
For $A\in\T(T_pM)$ and $Z\in T_pM$ one denotes the trace over the first two entries by
\begin{equation}\label{definition_c12}
c_{12}(A)(Z)=\sum_{i=1}^n A_{E_i E_i Z}.
\end{equation}
The space of quadratic invariants on $\T(T_pM)$ with respect to the $O(T_pM)$-representation is spanned by the three quadratic forms
\begin{eqnarray}
\|A \|^2 &=& \langle A,A \rangle\,, \label{def_torsionsnorm}\\
\langle A,\widehat{A} \rangle &=& \sum_{i,j,k=1}^n A_{E_i E_j E_k}A_{E_j E_i E_k}\,, \label{def_torsionsnormtwisted}\\
\|c_{12}(A)\|^2 &=& \sum_{i,j,k=1}^n A_{E_i E_i E_k}A_{E_j E_j E_k}\,.\label{def_torsionsspurnorm}
\end{eqnarray}
Here $\widehat{A}$ denotes the $(3,0)$-tensor obtained from $A$ by interchanging the first two slots, 
i.e.~$\widehat{A}_{XYZ} = A_{YXZ}$, for all tangent vectors $X,Y,Z$.

\begin{thm}\label{cartanklassifikation}
For $\dim(M)\ge 3$ one has the following decomposition of $\T(T_pM)$ into irreducible $O(T_pM)$-subrepresentations:
\[
\T(T_pM)\,=\,\T_1(T_pM)\,\oplus\, \T_2(T_pM)\,\oplus \,\T_3(T_pM).
\]
This decomposition is orthogonal with respect to $\langle\cdot,\cdot\rangle$, and it is given by
\begin{eqnarray*}
\T_1(T_pM)&=& \left\{ A\in \T(T_pM) \;\big|\; \exists V \mbox{ s.t. } \forall X,Y,Z:\, A_{XYZ}= 
\langle X,Y \rangle \langle V,Z \rangle-\langle X,Z \rangle\langle V,Y \rangle\right\}, \\
\T_2(T_pM)&=& \left\{ A\in \T(T_pM) \;\big|\; \forall X,Y,Z:\,A_{XYZ}=-A_{YXZ}\right\}, \\
\T_3(T_pM)&=& \left\{ A\in \T(T_pM) \;\big|\; \forall X,Y,Z:\,A_{XYZ}+ A_{YZX}+ A_{ZXY}=0\mbox{ and } c_{12}(A)(Z)=0\right\}.
\end{eqnarray*}
\noindent
For $\dim(M) = 2$ the $O(T_pM)$-representation 
\[
 \T(T_pM)\,=\,\T_1(T_pM)
\]
is irreducible.{\hfill$\boxbox$}
\end{thm}
The above theorem is just Thm.~3.1 from \cite{Tricerri}.
The connections whose torsion tensor is contained in $\T_1(T_pM)\cong T_pM$ are called {\it vectorial}.
Those whose torsion tensor is in $\T_2(T_pM)={\bigwedge}^3T^*_pM$ are called {\it totally anti-symmetric},
and those with torsion tensor in $\T_3(T_pM)$ are called {\it of Cartan type}.\medskip

\noindent
We note that any Cartan type torsion tensor $A\in \T_3(T_pM)$ is trace-free in any pair of entries, i.e.~for any $Z$ one has
\[
\sum_{i=1}^n A_{E_i E_i Z}=0, \qquad
\sum_{i=1}^n A_{E_i Z E_i}=0, \qquad
\sum_{i=1}^n A_{Z E_i E_i}=0.
\]
The second equality holds as $A\in \T(T_pM)$, and the third one follows from the cyclic identity $A_{XYZ}+ A_{YZX}+ A_{ZXY}=0$.

\begin{rem}\label{spurnullfuerT2undT3}
The invariant quadratic form given in (\ref{def_torsionsspurnorm}) has the null space $\T_2(T_pM)\oplus\T_3(T_pM)$.
More precisely, one has $A\in \T_2(T_pM)\oplus\T_3(T_pM)$ if and only if
$c_{12}(A)(Z)=0$ for any $Z\in T_pM$.{\hfill$\boxbox$}
\end{rem}

\begin{rem}\label{orthogonalTiforotherproduct}
The decomposition given in Theorem~\ref{cartanklassifikation} is orthogonal with respect to the bilinear form given in (\ref{def_torsionsnormtwisted}), i.e.~for $\alpha,\beta\in\{1,2,3\}$, $\alpha\ne\beta$, and $A_\alpha\in \T_\alpha(T_pM)$, $A_\beta\in \T_\beta(T_pM)$ one gets $\langle A_\alpha, \widehat{A}_\beta \rangle=0$. {\hfill$\boxbox$}
\end{rem}

\noindent
Varying the base point $p\in M$, the decomposition in Theorem~\ref{cartanklassifikation} is parallel with respect to the Levi-Civita connection $\nabla^g$ (induced on $(3,0)$-tensor fields).
And from Theorem~\ref{cartanklassifikation} one gets immediately:
\begin{cor}\label{connectionswithtorsionsclassified}
For any orthogonal connection $\nabla$ on some Riemannian manifold of dimension $n\ge 3$ 
there exist a vector field $V$, a $3$-form $T$ and a $(3,0)$-tensor field $S$ with $S_p\in \T_3(T_pM)$ for any $p\in M$ such that $\nabla_XY=\nabla^g_XY+A(X,Y)$ takes the form
\[
A(X,Y)=\langle X,Y \rangle V- \langle V,Y \rangle X + T(X,Y,\cdot)^\sharp + S(X,Y,\cdot)^\sharp ,
\]
where $T(X,Y,\cdot)^\sharp$ and $S(X,Y,\cdot)^\sharp$ are the unique vectors with
\bb
T(X,Y,Z)=\left\langle T(X,Y,\cdot)^\sharp,Z\right\rangle\mbox{ and }
S(X,Y,Z)=\left\langle S(X,Y,\cdot)^\sharp,Z\right\rangle\mbox{ for all }Z.
\label{sharp_def}
\ee
For any orthogonal connection these $V,T,S$ are unique.{\hfill$\boxbox$}
\end{cor}

\begin{lemma}
The scalar curvature of an orthogonal connection is given by
\[
R= R^g +2(n-1)\,\divergenz^{\nabla^g}(V) -(n-1)(n-2)\,\|V\|^2 -\|T\|^2+\tfrac{1}{2}\,\|S\|^2
\]
with $V,T,S$ as in Corollary~\ref{connectionswithtorsionsclassified}, and $\divergenz^{\nabla^g}(V)$ is the divergence of the vector field $V$ taken with respect to the Levi-Civita connection.
\label{Scalar_Norm}
\end{lemma}
\pf{ With the notations from (\ref{definition_c12})--(\ref{def_torsionsspurnorm}) we rewrite (\ref{Scalar_Allgemein}) as
\begin{equation}\label{eqn_R_general_morefancy}
R=R^g+2\,\sum_{i=1}^n c_{12}\left(\nabla^g_{E_i}A \right)(E_i) -\left\|c_{12}(A) \right\|^2 + \langle A, \widehat{A} \rangle.
\end{equation}
By Remark~\ref{spurnullfuerT2undT3} only the vectorial part of the torsion contributes to the $c_{12}$-terms, and therefore one gets
\begin{eqnarray}
\sum_{i=1}^n c_{12}\left(\nabla^g_{E_i}A \right)(E_i)&=& \sum_{i,j=1}^n 
\left(
\langle E_j,E_j \rangle\, \langle \nabla^g_{E_i}V,E_i\rangle- \langle\nabla^g_{E_i}V,E_j \rangle\,\langle E_i,E_j\rangle
\right)\nonumber\\
&=& (n-1)\,\divergenz^{\nabla^g}(V),\label{term_nablaspur}\\
\left\|c_{12}(A) \right\|^2 &=& \Big\|\sum_{j=1}^n c_{12}(A)(E_j)\,E_j\Big\|^2 \nonumber\\
&=& \Big\|\sum_{i,j=1}^n \left( \langle E_i,E_i \rangle\, \langle V,E_j\rangle\,E_j
- \langle V,E_i \rangle\,\langle E_i,E_j\rangle\,E_j
\right) \Big\|^2 \nonumber\\
&=& \Big\|\sum_{i=1}^n \left( V- \langle V,E_i \rangle\,E_i\right) \Big\|^2 \nonumber\\
&=& \|(n-1)\, V\|^2.\label{term_c12spur}
\end{eqnarray}
In order to compute the last term in (\ref{eqn_R_general_morefancy}) we decompose $A=A_1+A_2+A_3$ with $A_\alpha\in\T_\alpha(T_pM)$.
From Remark~\ref{orthogonalTiforotherproduct} we get 
\[
\langle A ,\widehat{A} \rangle=\sum_{\alpha=1}^3\langle A_\alpha ,\widehat{A}_\alpha \rangle.
\]
For the vectorial part we get
\begin{eqnarray}
\langle A_1,\widehat{A}_1\rangle  &=& \sum_{i,j,k=1}^n \big( \delta_{ij} \langle V,E_k\rangle -\delta_{ik}\langle V,E_j\rangle\big)\cdot 
\big( \delta_{ji} \langle V,E_k\rangle -\delta_{jk}\langle V,E_i\rangle\big)
\nonumber\\
&=& \sum_{i,j,k=1}^n \big(\delta_{ij}\, \langle V,E_k\rangle^2 - \delta_{ij}\delta_{jk}\, \langle V,E_k\rangle \langle V,E_i\rangle
-\delta_{ik}\delta_{ji}\, \langle V,E_j\rangle \langle V,E_k\rangle
+\delta_{ik}\delta_{jk}\, \langle V,E_j\rangle \langle V,E_i\rangle
\big) 
\nonumber\\
&=& (n-1)\,\|V\|^2
\end{eqnarray}
For the totally anti-symmetric part we get
\begin{equation}
\langle A_2,\widehat{A}_2\rangle \; = \sum_{i,j,k=1}^n T_{E_i E_j E_k}\,T_{E_j E_i E_k} 
= -\sum_{i,j,k=1}^n T_{E_i E_j E_k}\,T_{E_i E_j E_k}
\; =\;-\|T\|^2.
\end{equation}
Finally, for the Cartan-type part we get
\begin{eqnarray}
\langle A_3,\widehat{A}_3\rangle  &=& \sum_{i,j,k=1}^n S_{E_i E_j E_k}\,S_{E_j E_i E_k}\nonumber\\
&=& -\sum_{i,j,k=1}^n \big( S_{E_i E_j E_k}\,S_{´E_i E_k E_j}+S_{E_i E_j E_k}\,S_{E_k E_j E_i } \big) \label{cyclicschritt}\\
&=& \sum_{i,j,k=1}^n  S_{E_i E_j E_k}\,S_{´E_i E_j E_k}- \sum_{i,j,k=1}^nS_{E_i E_k E_j}\,S_{E_k E_i E_j }  \label{antisymmschritt}\\
&=& \tfrac12\,\|S\|^2, \label{cartantype-anteil}
\end{eqnarray}
where (\ref{cyclicschritt}) is due to the cyclic identity for $S$, (\ref{antisymmschritt}) follows from the anti-symmetry in the last two entries, and 
$ \sum_{i,j,k=1}^nS_{E_i E_k E_j}\,S_{E_k E_i E_j }= \langle A_3,\widehat{A}_3\rangle $ implies (\ref{cartantype-anteil}). 
Plugging (\ref{term_nablaspur})--(\ref{cartantype-anteil}) into (\ref{eqn_R_general_morefancy}) finishes the proof.
{\hfill$\boxbox$}
}

\begin{cor}\label{Einstein_Cartan_Hilbert}
Let $M$ be a closed manifold of dimension $n\ge 3$ with Riemannian metric $g$ and orthogonal connection $\nabla$.
Let $\dvol$ denote the Riemannian volume measure taken with respect to $g$.
Then the Einstein-Cartan-Hilbert functional is 
\[
 \int_M R \dvol = \int_M R^g \dvol -\,(n-1)(n-2)\,\int_M \|V\|^2\dvol -\int_M \|T\|^2 \dvol+\tfrac{1}{2}\,\int_M \|S\|^2\dvol.
\]
\end{cor}
\noindent
Considering variations over all Riemannian metrics, for which the volume $\vol_g(M)$ stays fixed, and all orthogonal connections (i.e.~over all torsion tensors), we get that $(M,g,\nabla)$ is a critical point of the Einstein-Cartan-Hilbert functional if and only if  $(M,g)$ is an Einstein manifold and $\nabla=\nabla^g$ is the Levi-Civita connection (i.e.~$V\equiv 0$, $T\equiv 0$ and $S\equiv 0$).

\noindent 
For an in depth treatment of the physical consequences of Einstein-Cartan-Hilbert theory
in Lorentzian geometry we refer to the classical review \cite{HHKN76} and the more 
recent overview  \cite{S02} and references therein.

\section{Curvature calculations in case of totally anti-symmetric torsion
in four dimensions}\label{section2}
Let us collect now some equalities involving curvature tensors and the 
totally anti-symmetric torsion. To keep the main part of this paper as readable
as possible the proofs of the following theorems and lemmata have been 
allocated to the appendix.
 
\noindent 
We consider a $4$-dimensional manifold $M$ equipped with a Riemannian metric $g$.
Let $\nabla^{g}$ denote the Levi-Civita connection on the tangent bundle.
We fix some $3$-form $T$ on $M$ and some $s\in\R$, and we are studying the connection $\nabla$ which is given by
\begin{equation}
\label{vectortorsionconnection}
\nabla_X Y = \nabla^g_X Y + s \, T(X,Y,\cdot)^\sharp
\end{equation}
for any vector fields $X$ and $Y$ on $M$ and $T(X,Y,\cdot)^\sharp$ is defined as in (\ref{sharp_def}). 
Hence $\nabla$ is an orthonormal connection with totally anti-symmetric torsion.
\subsection{Pointwise equalities}
In the following we want to express some curvature quantities for $\nabla$ in terms of $T$ and curvature quantities for $\nabla^g$. For the Riemann curvature
and the scalar curvature we will calculate the norms explicitly
in terms of $T$, the Levi-Civita connection  and its curvatures.
The norm of the Ricci curvature is given in the appendix.
\medskip

\noindent
As in section~\ref{section1} we fix some point $p\in M$, and we extend any tangent vectors $X,Y,Z,W\in T_p M$ to vector fields again denoted by $X,Y,Z,W$ being synchronous in $p$.
Hence we obtain from (\ref{Riemann_Torsion_Allgemein}) the 
Riemann curvature of $\nabla$
\begin{eqnarray}
\left\langle \Riem(X,Y)Z,W \right\rangle&=&\left\langle \Riem^g(X,Y)Z,W \right\rangle + 
s\; \left( \left(\nabla^g_XT \right)(Y,Z,W)-\left(\nabla^g_YT \right)(X,Z,W)\right) \nonumber \\
&&\quad +s^2\,\left( T\left(X,T(Y,Z,\cdot)^\sharp,W \right) -T\left(Y,T(X,Z,\cdot)^\sharp,W \right) \right)
\nonumber\\
&=&\left\langle \Riem^g(X,Y)Z,W \right\rangle + s \; \left( \left(\nabla^g_XT \right)(Y,Z,W)-\left(\nabla^g_YT \right)(X,Z,W)\right) \nonumber \\
&&\quad +s^2\,\left(\left\langle T(X,Z,\cdot)^\sharp,T(Y,W,\cdot)^\sharp \right\rangle - \left\langle T(Y,Z,\cdot)^\sharp,T(X,W,\cdot)^\sharp\right\rangle
\right).
\label{Riemann_identity}
\end{eqnarray}
We used the identity 
$T(X,T(Y,Z,\cdot)^\sharp,W)=-\left\langle T(X,W,\cdot)^\sharp,T(Y,Z;\cdot)^\sharp \right\rangle$, which follows from (\ref{sharp_def}).\medskip

\noindent
From (\ref{Ricci_Torsion_Allgemein}) we conclude that the Ricci curvature of 
$\nabla$ for any orthonormal,
synchronous frame $E_1,\ldots,E_n$  defined on some neighbourhood of $p$  is 
\begin{eqnarray}
\ric(X,Y)&=& \ric^g(X,Y) +s \;\sum_{i=1}^n \left(\nabla^g_{E_i}T \right)(X,Y,E_i)
-s^2\;\sum_{i=1}^n\left\langle T(E_i,X,\cdot)^\sharp,T(E_i,Y,\cdot)^\sharp \right\rangle.
\label{Ricci_identity}
\end{eqnarray}
This formula shows that in general the Ricci curvature is not symmetric in $X$ and $Y$.
\medskip

\noindent
From Lemma \ref{Scalar_Norm}
we get for the scalar curvature $R$ of $\nabla$ that
\begin{eqnarray}
R =R^g-s^2\, \left\|T \right\|^2.
\label{Scalar_identity}
\end{eqnarray}

\noindent
Next, we are aiming at finding an expression for the norm of the Riemann tensor.
As the vector fields $X,Y,Z,W$ are synchronous in $p$ we get for the differential $dT$ and the codifferential $\delta T$ of the $3$-form $T$:
\begin{align}
dT(X,Y,Z,W)&= \left(\nabla^g_X T\right)(Y,Z,W) -\left(\nabla^g_Y T\right)(X,Z,W)+\left(\nabla^g_Z T\right)(X,Y,W)-\left(\nabla^g_W T\right)(X,Y,Z),\label{exterior_differential}\\
\delta T(X,Y)&=-\sum_{i=1}^n\left(\nabla^g_{E_i} T\right)(X,Y,E_i).
\label{exterior_codifferential}
\end{align}
\noindent
We define the $(4,0)$-tensors $\riem$ and $\riem^g$ by
\begin{equation}\label{def_40riem}
\riem^{(g)}(X,Y,Z,W)=\left\langle \Riem^{(g)} (X,Y)Z,W \right\rangle.
\end{equation}
We decompose the Riemann tensor into its symmetric 
and  anti-symmetric component
\bb
\riem(X,Y,Z,W) = \riem^S(X,Y,Z,W) + \riem^A(X,Y,Z,W).
\label{Riem_decomp}
\ee
The symmetric part of $\riem$ is 
\bb
\riem^S(X,Y,Z,W)=\tfrac{1}{2}\left(\riem(X,Y,Z,W)+ \riem(Z,W,X,Y)\right)
\eee
and the anti-symmetric part of $\riem$ is then given by
\bb
\riem^A(X,Y,Z,W)=\tfrac{1}{2}\left(\riem(X,Y,Z,W)- \riem(Z,W,X,Y)\right).\label{antisymm_Riemann}
\eee
\noindent
Since $\riem^S$ and $\riem^A$ are orthogonal with respect to the scalar product of
$(4,0)$-tensors, as defined in (\ref{deftensorprod}), we find
\bb
\|\Riem \|^2 = \| \riem  \|^2 = \| \riem^S  \|^2 + \| \riem^A  \|^2
\label{Riemnormsqared}
\ee
We also decompose the Ricci curvature 
into its symmetric and its anti-symmetric components 
\begin{eqnarray}
\ric(X,Y) &=&\ric^S(X,Y) + \ric^A(X,Y),
\label{Ric_decomp}\\
\mbox{with }\quad
\ric^S(X,Y)&=&\tfrac{1}{2}\,\left(\ric(X,Y)+\ric(Y,X) \right)
\nonumber\\
\mbox{and  }\quad
\ric^A(X,Y)&=&\tfrac{1}{2}\,\left(\ric(X,Y)-\ric(Y,X) \right).\nonumber
\end{eqnarray}

\noindent
Now we  give an explicit formula
for $\|\riem\|^2$ in the case of $M$ being $4$-dimensional.
\begin{thm}\label{theoremriem}
Let $M$ be a $4$-dimensional manifold with Riemannian metric $g$ and 
connection $\nabla$ as given in (\ref{vectortorsionconnection}).
Then the norm of the Riemann tensor of $\nabla$ is given by
\bb
\left\|\Riem \right\|^2 = \left\|\riem^g \right\|^2 +  \tfrac{1}{3} s^4\,  \left\| T \right\|^4 
+ \tfrac{1}{4} s^2 \,  \left\|dT \right\|^2 
- \tfrac{1}{3}s^2 \,R^g\,\left\|T\right\|^2
+4 s^2 \,\BB(T)  + \left\|\riem^A \right\|^2 
\label{Ricnormsqared}
\ee
with 
\bb
\BB(T) = \sum_{i,j,k} \ric^g(E_i,E_k)\, 
\left\langle T(E_i,E_j,\cdot)^\sharp,T(E_j,E_k,\cdot)^\sharp \right\rangle
+\tfrac{1}{4}\,R^g\,\left\| T\right\|^2
\label{BBT}
\eee
\end{thm}
\pf{See Appendix.}
\medskip

\noindent
We notice that the term $\BB(T)$ couples the torsion to the Ricci curvature, and the term $\left\|\riem^A \right\|^2 $
is being computed in Lemma~\ref{riemAsquare}.

\subsection{Integral formulas}

In this section $(M,g)$ will be a closed, $4$-dimensional Riemannian manifold.
We will exploit the topological invariance of the Euler characteristic to
deduce integral formulas for $3$-forms defined on $M$.

\begin{Def}
Let $\nabla$ be an orthogonal connection on $M$ 
and let $\Riem_{ij}(X,Y):= \langle \Riem(E_i,E_j)X,Y\rangle$ be its
curvature $2$-form defined by equation  (\ref{Riemann_Torsion_Allgemein}). 
Define the $4$-form
\[
{\bf K} = \tfrac{1}{32 \pi^2 } \sum_{i,j,k,l=1}^4 \epsilon_{ijkl} \Riem_{ij}\wedge
\Riem_{kl},
\]
where $\epsilon_{ijkl}$ is the totally anti-symmetric tensor with normalisation 
$\epsilon_{1234} = +1$.
\end{Def}
One obtains the classical result for the interplay between the topological invariant
Euler characteristic $\chi(M)$ and the curvature $2$-form of $\nabla$:
\begin{thm}
The Euler characteristic of $M$ is 
\[
\chi(M) = \int_M {\bf K}.
\] 
\end{thm} 
\pf{
For a proof of this theorem we refer to \cite[Vol.~II, Chap.~XII, Thm.~5.1]{KN69}.
\hfill$\boxbox$
}
\medskip

\noindent
In four dimensions the Euler characteristic can be expressed in a particularly
convenient form in terms of squares of the Riemann, Ricci and 
scalar curvature of $\nabla$.

\begin{thm}
Let  $\nabla$ an orthogonal connection  on $M$ and  
let $\riem= \riem^S + \riem^A$, $\ric= \ric^S + \ric^A$ and $R$ be the 
Riemann curvature, the
Ricci curvature and the scalar curvature of $\nabla$ decomposed into
their symmetric and anti-symmetric parts according to (\ref{Riem_decomp})
and (\ref{Ric_decomp}).  
Then the Euler characteristic $\chi(M)$ is 
\[
\chi(M) = \tfrac{1}{8 \pi^2} \, \int_M \left(R^2 - 4\, \| \ric^S \|^2 
+ 4\, \| \ric^A \|^2 + \| \riem^S \|^2 -  \| \riem^A \|^2 \right)\, dvol. 
\]
\label{Eulerallg}
\end{thm}
\pf{See Appendix.}
\medskip

\noindent
The classical result for the Euler characteristic in terms of the curvatures of the Levi-Civita connection is due to Berger \cite{Berger}, 
and it follows immediately from the above theorem: 
\begin{cor}\label{Eulerspezial}
Let $M$ be a $4$-dimensional manifold with Riemannian metric $g$ and 
Levi-Civita connection $\nabla^g$.
Then the Euler characteristic $\chi(M)$ is given by
\[
\chi(M) = \tfrac{1}{8 \pi^2} \, \int_M \left((R^g)^2 - 4\, \| \ric^g \|^2 
 + \| \riem^g \|^2 \right)\, dvol 
\]
\end{cor}
\pf{
For $\nabla^g$ we have $\riem^S = \riem^g$, $\ric^S = \ric^g$  and hence  $\riem^A\equiv 0$ and $\ric^A\equiv 0$.
\hfill$\boxbox$
}

\noindent
The fact that the Euler characteristic does not depend on the connection
allows us to deduce a useful integral formula for $3$-forms
on closed Riemannian $4$-manifolds.

\begin{lemma}\label{Eulervanish}
Let $M$ be a closed $4$-dimensional manifold with Riemannian metric $g$  
and $T$ any $3$-form on $M$. 
Let  $R^g$ denote the scalar curvature of the Levi-Civita connection $\nabla^g$.  
Then
\bb
\int_M   4 \| \delta T \|^2  \, dvol 
= \int_M \left(    \tfrac{1}{3} \, R^g \| T \|^2  - \tfrac{1}{4} \|dT\|^2  + 4 B(T)
+ \frac{1}{s^2} \| \riem^A \|^2\right) \, dvol
\label{vanish}
\ee
with $\BB(T)$ as defined in Theorem \ref{theoremriem} and $\riem^A$ 
is the anti-symmetric component of the Riemann curvature of $\nabla$ with
$s \,T$ as torsion $3$-form.
\label{intzero}
\end{lemma}
\pf{ See Appendix.}

\section{Dirac operators associated to orthogonal connections}\label{section3}

In this section we consider an $n$-dimensional oriented Riemannian manifold $(M,g)$ equipped with some spin structure.
Let $\nabla$ be an orthogonal connection given as in (\ref{def_Zusammenhang_mit_Torsion})--(\ref{A_schiefadjungiert}).
Then the connection $\nabla$ acting on vector fields induces a connection acting on spinor fields.
Next, we will briefly discuss the construction of this connection (compare \cite[p.~17f]{AgricolaSrni}, or see \cite[Chap.~II.4]{Lawson} for more details).
Again, we write $\nabla_XY=\nabla^g_XY+A(X,Y)$ with the Levi-Civita connection $\nabla^g$.
For any $X\in T_pM$ the endomorphism $A(X,\cdot)$ is skew-adjoint and hence it is an element of $\mathfrak{so}(T_pM)\cong\bigwedge^2T_pM$, we can express it as
\begin{equation}\label{torsionendo_in_terms_of_son}
A(X,\cdot) =\sum_{i< j} \alpha_{ij} \,E_i\wedge E_j.
\end{equation}
Here $E_i\wedge E_j$ is meant as the endomorphism of $ T_pM$ defined by $E_i\wedge E_j(Z)= \langle E_i,Z\rangle E_j-\langle E_j,Z\rangle E_i$.
For any $X\in T_pM$ one determines the coefficients in (\ref{torsionendo_in_terms_of_son}) by
\begin{equation}\label{determination_of_alphas}
\alpha_{ij}=\langle A(X,E_i), E_j \rangle= A_{X E_i E_j}.
\end{equation}
Each $E_i\wedge E_j$ lifts to $\frac12 E_i\cdot E_j$ in $\mathfrak{spin}(n)$, and the spinor connection induced by $\nabla$ is locally given by
\begin{equation}\label{spintorsionconnection}
\nabla_X\psi = \nabla^g_X\psi+\tfrac12 \sum_{i< j} \alpha_{ij} \, E_i\cdot E_j \cdot \psi 
 = \nabla^g_X\psi+\tfrac12 \sum_{i< j} A_{X E_i E_j} \, E_i\cdot E_j \cdot \psi. 
\end{equation}
\begin{rem}
The connection given by (\ref{spintorsionconnection}) is compatible with the metric on spinors and with Clifford multiplication (see e.g.~\cite[Lemma~2.1]{AgricolaSrni}).
\end{rem}
\begin{rem}\label{spinorconnection_totallyanti}
For totally anti-symmetric torsion, given by a $3$-form $T$ as in Corollary~\ref{connectionswithtorsionsclassified}, one can rewrite (\ref{spintorsionconnection}) as
\[
\nabla_X \psi = \nabla^{g}_X \psi + \frac{1}{2} (X \lrcorner T) \cdot \psi,
\]
where $X \lrcorner T$ is the $2$-form defined by $X \lrcorner T(X,Z)=T(X,Y,Z).$
We recall that a $k$-form $\omega\in \bigwedge^kT_pM$, given as
$\omega=\sum\limits_{i_1,\ldots,i_k}\omega_{i_1,\ldots,i_k} \,{E_{i_1}}^\flat\wedge\ldots\wedge {E_{i_k}}^\flat$, acts on the spinor space as $\omega\cdot\psi= \sum\limits_{i_1,\ldots,i_k}\omega_{i_1,\ldots,i_k}\, E_{i_1}\cdot \ldots\cdot E_{i_k}\cdot \psi$.
\end{rem}
\begin{rem}
Not any connection on spinor fields is induced by an orthogonal connection on tangent vector fields.
For example, for  the connection
$
\nabla_X\psi = \nabla^g_X\psi + X\cdot \psi
$
the endomorphism $\alpha$ of spinors defined by
\begin{equation}\label{tetraden}
\alpha(\psi)=\nabla_X\left(Y\cdot\psi\right)-Y\cdot\left(\nabla_X \psi\right)
\end{equation}
is given by multiplication by the Clifford element $\nabla_X^g Y+X\cdot Y-Y\cdot X$, which does not equal to the Clifford multiplication by any tangent vector. This consideration applies in any dimension $n\ge 2$.
\end{rem}
\begin{rem}
If we assume that for a spinor connection $\nabla$ for any vector fields $X,Y$ the endomorphism $\alpha$ defined in (\ref{tetraden}) is the Clifford multiplication by a tangent vector $V_{X,Y}$, i.e.~$\alpha(\psi)=V_{X,Y}\cdot \psi$ for all spinors $\psi$, then it can be shown than the assignment $\nabla_XY=V_{X,Y}$ defines an orthogonal connection on tangent vector fields such that the spinor connection is compatible with the Clifford multiplication. 
In that case, physics literature occasionally refers to (\ref{tetraden}) as the {\it tetrad postulate}.
\end{rem}

\noindent
The Dirac operator associated to the spinor connection from (\ref{spintorsionconnection}) is defined as
\begin{eqnarray}
D\psi &=& \sum_{i=1}^n E_i\cdot \nabla_{E_i}\psi \nonumber\\
&=&  D^g\psi +\tfrac12\sum_{i=1}^n \sum_{j<k} A_{E_i E_j E_k} \, E_i\cdot E_j \cdot E_k \cdot \psi
\nonumber\\
&=& D^g\psi +\tfrac14\sum_{i,j,k=1}^n  A_{E_i E_j E_k} \, E_i\cdot E_j \cdot E_k \cdot \psi, \label{def_torsiondirac}
\end{eqnarray}
where $D^g$ is the Dirac operator induced by the Levi-Civita connection.\medskip

\noindent
The next theorem tells us when the Dirac operator $D$ is formally selfadjoint (i.e.~symmetric on the space of compactly supported smooth spinor fields as domain), it is provided as Satz 2 in \cite{FriedrichSulanke}:
\begin{thm}
The Dirac operator $D$ is formally selfadjoint if and only if the divergence of $\nabla$ coincides with the divergence of $\nabla^g$, i.e.~for any vector field $Z$ one has
\begin{equation}\label{divergenz_gleichheit}
\sum_{i=1}^n \langle \nabla_{E_i} Z,E_i \rangle =
\sum_{i=1}^n \langle \nabla^g_{E_i} Z,E_i\rangle
\end{equation}
in any point $p$ and for any orthonormal basis $E_1,\cdots, E_n$ of $T_pM$. {\hfill$\boxbox$}
\end{thm}
Taking the specific form of $\nabla_XY=\nabla^g_XY+A(X,Y)$ into account, we see that (\ref{divergenz_gleichheit}) is equivalent to 
\[
c_{12}(A)(Z) = \sum_{i=1}^n \langle A(E_i,E_i),Z \rangle=-\sum_{i=1}^n \langle A(E_i,Z),E_i\rangle =0.
\]
Hence, we can conclude from Remark~\ref{spurnullfuerT2undT3}:
\begin{cor}\label{vectornull_dannselbstadjungiert}
The Dirac operator $D$ associated to an orthogonal connection is formally selfadjoint if and only if the $(3,0)$--torsion tensor $A$ does not have any vectorial compontent, i.e.~one has
\[
A_p\in\T_2(T_pM)\oplus\T_3(T_pM)
\]
in any point $p\in M$ {\hfill$\boxbox$}
\end{cor}

\noindent
The next lemma states that for the Dirac operator the Cartan type component of the torsion is invisible:

\begin{lemma}\label{Cartan_type_egal}
On a Riemannian spin manifold we consider some vector field $V$, some $3$-form $T$ and some $(3,0)$-tensor field $S$ with $S_p\in\T_3(T_pM)$ for any $p\in M$.
Let $\nabla_1$ and $\nabla_2$ be the orthogonal connections determined by
\begin{align*}
A_1(X,Y) & = \langle X,Y \rangle V- \langle V,Y \rangle X + T(X,Y,\cdot)^\sharp + S(X,Y,\cdot)^\sharp \quad\mbox{ and}\\
A_2(X,Y) & =\langle X,Y \rangle V- \langle V,Y \rangle X + T(X,Y,\cdot)^\sharp,
\end{align*}
respectively (compare Corollary~\ref{connectionswithtorsionsclassified}).
Denote the associated Dirac operators by $D_1$ and $D_2$.
Then, for any spinor field $\psi$ one has 
\[
D_1 \psi =D_2\psi.
\]
\end{lemma}
{\pf  
By (\ref{def_torsiondirac}) the difference of the two Dirac operators is
\begin{equation}\label{difference_Dirac_operators}
D_1\psi -D_2\psi =\tfrac14\sum_{i,j,k=1}^n  S_{E_i E_j E_k} \, E_i\cdot E_j \cdot E_k \cdot \psi.
\end{equation}
We use the cyclic identity for $S$, the fact that $S$ is trace-free in any pair of entries and the Clifford relations $E_i\cdot E_j=-E_j\cdot E_i$ for $i\ne j$ as well, in order to obtain:
\begin{align*}
\sum_{i,j,k=1}^n  S_{E_i E_j E_k} \, E_i\cdot E_j \cdot E_k
&= -\sum_{i,j,k=1}^n  S_{E_j E_k E_i} \, E_i\cdot E_j \cdot E_k -\sum_{i,j,k=1}^n  S_{E_k E_i E_j} \, E_i\cdot E_j \cdot E_k \\
&= -\sum_{i,j,k=1}^n  S_{E_j E_k E_i} \, E_j\cdot E_k \cdot E_i -\sum_{i,j,k=1}^n  S_{E_k E_i E_j} \, E_k\cdot E_i \cdot E_j \\
&= -2\,\sum_{i,j,k=1}^n  S_{E_i E_j E_k} \, E_i\cdot E_j \cdot E_k.
\end{align*}
Therefore we get $\sum\limits_{i,j,k=1}^n  S_{E_i E_j E_k} E_i\cdot E_j \cdot E_k=0$, and the right hand side of (\ref{difference_Dirac_operators}) is zero.
{\hfill$\boxbox$}
}\medskip

\noindent
One should note that the above lemma applies pointwise.

\begin{rem}
In the Lorentzian case it is known that torsion of Cartan type does not 
contribute to the Dirac action under the integral \cite[Chap.~2.3]{S02}. 
It is also known that the Dirac action is not real if the torsion has a 
non-vanishing vectorial component \cite[Chap.~11.6]{GS87}.
Therefore only totally anti-symmetric torsion is considered to couple 
to fermions reasonably. 
\end{rem}

\noindent
The spinor connection in (\ref{spintorsionconnection}) is the connection which is induced by the tangent vector connection $\nabla$ given in (\ref{def_Zusammenhang_mit_Torsion}), hence one expects that their curvatures are related.
Let $(E_1,\ldots,E_n)$ be an arbitrary local orthonomal frame.
For $i,j$ the curvature endomorphism w.r.t.~this frame is defined as
\bb
\Omega_{ij}\psi =\nabla_{E_i}\nabla_{E_j} \psi-\nabla_{E_j}\nabla_{E_i}\psi -\nabla_{[E_i,E_j]}\psi.
\eee
These curvature endomorphisms for spinors can be naturally determined by the Riemann tensor for tangent vectors, compare with formula (4.37) in  Theorem 4.15 of \cite[Chap.~II]{Lawson}.
\begin{lemma}
\label{curvaturetrace}
For the spinor connection $\nabla$ defined in (\ref{spintorsionconnection}) the curvature endomorphisms in $p$ are given by
\[
\Omega_{ij}\psi=\tfrac{1}{4}\,\sum_{a,b} \left\langle\Riem(E_i,E_j)E_a,E_b\right\rangle\,E_a\cdot E_b\cdot\psi
\]
with Riemann tensor as defined in (\ref{Riemann_Torsion_Allgemein}).
\hfill$\boxbox$
\end{lemma}

\begin{cor}\label{Omega-trace-Riem}
Let $\tr$ denote the trace over the spinor space over some footpoint $p$.
Then one has
\[
\sum_{i,j=1}^n\,\tr\left( \Omega_{ij}\Omega_{ij} \right)= -\tfrac{1}{8}\cdot 2^{[n/2]}\cdot\left\|\Riem \right\|^2
\]
where $\Riem$ is the Riemann tensor of the vector connection $\nabla$.
\end{cor}
\pf{
The spinor space has dimension $2^{[n/2]}$.
If $a\ne b$ and $c\ne d$  the Clifford relations imply
\[
\tr\left(E_a\cdot E_b\cdot E_c\cdot E_d \right) = 2^{[n/2]}\left(\delta_{bc}\delta_{ad}-\delta_{bd}\delta_{ac} \right).
\]
From Lemma~\ref{curvaturetrace} we derive
\begin{align*}
\sum_{i,j}\,\tr\left( \Omega_{ij}\Omega_{ij} \right)
&=\tfrac{1}{16}\,\sum_{i,j}\, \sum_{a\ne b}\, \sum_{c\ne d}
\left\langle\Riem(E_i,E_j)E_a,E_b\right\rangle\,\left\langle\Riem(E_i,E_j)E_c,E_d\right\rangle\,
\tr\left(E_a\cdot E_b\cdot E_c\cdot E_d \right)
\\
&=\tfrac{1}{16}\cdot 2^{[n/2]}\,\sum_{i,j}\, \sum_{a\ne b}\,\big(
\left\langle\Riem(E_i,E_j)E_a,E_b\right\rangle\,\left\langle\Riem(E_i,E_j)E_b,E_a\right\rangle\\
&\qquad\qquad\qquad\qquad\qquad
- \left\langle\Riem(E_i,E_j)E_a,E_b\right\rangle\,\left\langle\Riem(E_i,E_j)E_a,E_b\right\rangle
\big)\\
&=-\tfrac{1}{8}\cdot 2^{[n/2]}\,\sum_{i,j}\, \sum_{a\ne b}\,\left(\left\langle\Riem(E_i,E_j)E_a,E_b\right\rangle\right)^2
\\
&= -\tfrac{1}{8}\cdot 2^{[n/2]}\cdot\left\|\Riem \right\|^2,
\end{align*}
where we have used the anti-symmetry of $\Riem(E_i,E_j)E_a,E_b$ in the indices $a$ and $b$, which holds due to (\ref{Riemann_Torsion_antisymm_in_34_indices}).
\hfill$\boxbox$}

\section{Commutative geometries and the spectral action principle}\label{section4}
In this section we want to discuss torsion connections within the framework of Connes' noncommutative geometry (see \cite{Connes94}).
Let $M$ be a closed Riemannian manifold of dimension $n$ with some fixed spin structure.
We denote the algebra of smooth functions by $\A=C^\infty(M)$, and we denote the Hilbert space of square integrable spinor fields by $\HH$.
The Dirac operator $D^g$ associated to the Levi-Civita connection is a selfadjoint operator in $\HH$.
The triple $(\A,\HH,D^g)$ forms a {\it canonical spectral triple}, and it satisfies all {\it axioms for commutative geometry} (see \cite{Connes96}, or \cite{CostaRica} for more details).\medskip

\noindent
Now, let $\nabla$ be an orthogonal connection on the tangent bundle of $M$, and let $D$ denote the associated Dirac operator.
By Corollary~\ref{vectornull_dannselbstadjungiert} we know that $D$ is symmetric if and only if the vectorial component of the torsion of $\nabla$ is zero.
In that case $D$ is the sum of a selfadjoint operator and a bounded symmetric one, and thus selfadjoint.
We notice that $D$ has the same principal symbol and the same Weyl asymptotics as $D^g$.
Furthermore, we note that any natural algebraic structure on the spinor space such as a real structure or the Clifford multiplication with the volume element are parallel with respect to any spinor connection which is induced by an orthogonal connection on the tangent bundle.
Therefore $D$ commutes or anti-commutes with such a structure exactly if $D^g$ does.
Following the details of the proof of \cite[Thm.~11.1]{CostaRica} we see that these observations suffice to verify all axioms for commutative geometry
and we conclude:
\begin{lemma}\label{commutative_geometry}
If the vectorial component of the torsion of $\nabla$ is zero,
the spectral triple $(\A,\HH,D)$ satisfies the axioms for commutative geometry.
\hfill$\boxbox$
\end{lemma}
Connes' Reconstruction Theorem (conjectured in \cite{Connes96}, proved in \cite{Connes08}) states that, given a commutative geometry $(\A,\HH,D)$, one can construct a differentiable spin manifold such that $\A$ coincides with the smooth functions on it. 
Then, from the data $(\A,\HH,D)$ one also gets a Riemannian metric and one obtains that $\HH$ is isomorphic to the square integrable spinor fields (for some spin structure) (see \cite[Th\'eor\`eme 6]{Connes95}), and the natural Dirac operator we can always construct is the one induced by the Levi-Civita connection.\medskip

\noindent
In the above situation of Lemma~\ref{commutative_geometry}, one can algebraically recover the totally anti-symmetric torsion component from the spectral triple $(\A,\HH,D)$ if the underlying manifold $M$ has even dimension.
This can be done by considering the endomorphism of spinors given as difference of $D$ and the Levi-Civita Dirac operator $D^g$, see (\ref{def_torsiondirac}).
In even dimensions the complex Clifford algebra and the space of endomorphisms of the spinor space are identical (see~\cite{Friedrich1997}, Proposition on p.~13), and hence the endomorphism $D-D^g$ can be uniquely determined as a $3$-form.
In odd dimensions this argument does not apply, as one easily sees by noticing that e.g.~in the $3$-dimensional case the volume form acts as multiple of the identity on the spinor space.\medskip

\begin{figure}[h]
 \begin{pspicture}(-7,-2.1)(4,2.8)
\psline[fillcolor=lightgray,fillstyle=solid,linewidth=0,linecolor=lightgray](-1.3,-1)(-1.3,2)(4,2)(4,-1)
\psline[linewidth=0.03cm,linestyle=dotted](1.8,1.8)(0,0)
\psline[linewidth=.03cm,linecolor=blue](-1.2,-1)(-1.2,2)
\psline[linewidth=.03cm,linecolor=blue](-.6,-1)(-.6,-.7)
\psline[linewidth=.03cm,linecolor=blue](-.6,-.5)(-.6,2)
\psline[linewidth=.03cm,linecolor=blue](.6,-1)(.6,2)
\psline[linewidth=.03cm,linecolor=blue](1.2,-1)(1.2,2)
\psline[linewidth=.03cm,linecolor=blue](1.8,-1)(1.8,2)
\psline[linewidth=.03cm,linecolor=blue](2.4,-1)(2.4,2)
\psline[linewidth=.03cm,linecolor=blue](3,-1)(3,2)
\psline[linewidth=.03cm,linecolor=blue](3.6,-1)(3.6,2)
\psline[linewidth=0.05cm]{->}(-1.3,0)(4,0)
\psline[linewidth=0.05cm]{->}(0,-1)(0,2)
\psline[linewidth=0.05cm]{->}(0,0)(-1.5,-1.5)
\rput(3.8,-.25){$T$}
\rput(-.25,1.8){$S$}
\rput(-1.05,-1.4){$V$}
 \end{pspicture}
\caption{Cartan-type component $S$ of torsion is invisible for Dirac operator (gauge freedom).}
\label{figure_torsion_types}
\end{figure}
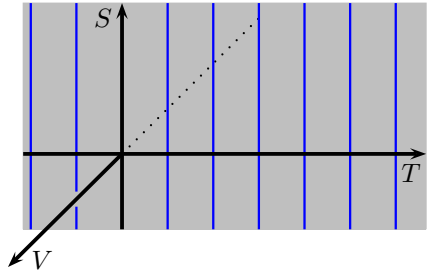

\noindent
By Lemma~\ref{Cartan_type_egal} the Cartan-type component of the torsion is invisible for the Dirac operator $D$, and therefore it cannot be recovered from $(\A,\HH,D)$.
This can be interpreted as some sort of gauge freedom, which is schematically illustrated in Figure~\ref{figure_torsion_types}.

\begin{rem}
For some even-dimensional compact Riemannian manifold $M$ with a given spin structure we fix $\A=C^\infty(M)$ and $\HH$ the space of square integrable spinor fields.
The above considerations show that one has a family of Dirac operators parametrized by the $3$-forms $T\in \Omega^3(M)$ such that the associated spectral triples are pairwise disctinct commutative geometries. 
Notice that all these Dirac operators are in the same $K$-homology class since they all have the same principal symbol.
We leave it open how big the class of first order operators $D$ in $\HH$ is for which
 $(\A,\HH,D)$ forms a commutative geometry. 
\end{rem}

\begin{rem}
For spectral triples of odd $KO$-dimension  it has recently been 
shown in \cite[Prop.~1.2]{SZ10} that one can modify the Dirac operator by adding a term induced
by a selfadjoint element of the algebra $\A$ and still finds the axioms of spectral triples satisfied.
In the case of $\A=C^\infty(M)$ and $KO$-dimensions  3 or 7 modulo 8, i.e.~$M$ is of dimension 3 or 7 modulo 8,
this modification is realised by adding a real-valued function
$\Phi \in \A$ to the Dirac operator, see  \cite[Rem.~1.3]{SZ10}. 
\end{rem}

\noindent
In the following we will only consider orthogonal connections $\nabla$ with zero vectorial component to ensure selfadjointness of the induced Dirac operator $D$.
For the computation of the Chamseddine-Connes spectral action (see \cite{ConnesChamseddine1}) we need the Seeley-deWitt coefficients $a_{2k}(D^2)$ of the heat trace asymptotics \cite{Gilkey95}
\bb
{\rm Tr} \left( e^{-t \, D^2} \right) \sim  \sum_{k\geq 0} t^{k- n/2} a_{2k} (D^2) \quad \mbox{ as }t\to 0.
\nonumber
\ee 
\begin{prop}
The first two Seeley-deWitt coefficients are
\begin{align*}
a_0(D^2)&= \frac{1}{(4\pi)^{n/2}} \;  2^{[n/2]}\;\int_M dvol, \\
a_2(D^2)&= \frac{1}{(4\pi)^{n/2}} \;  2^{[n/2]}\;\int_M \left(\frac{3}{4}\|T\|^2-\frac{1}{12} R^g \right) dvol.
\end{align*}
\end{prop}
\pf{ 
By Lemma~\ref{Cartan_type_egal} we can assume without loss of generality that the Cartan-type component of the torsion vanishes.
The orthogonal connection $\nabla$ is given by
$
\nabla_XY= \nabla^g_XY +T(X,Y,\cdot)^\sharp.
$
Adapting \cite[Thm.~6.2]{AgricolaFriedrich} into our notation we get the Bochner formula
\begin{equation}\label{TorsionsBochner}
D^2=\Delta +\frac{3}{2}dT +\frac{1}{4}R^g -\frac{3}{4}\|T \|^2
\end{equation}
where $\Delta$ is the Laplacian associated to the spin connection
\bb
\widetilde{\nabla}_X \psi = \nabla^{g}_X \psi + \frac{3}{2} (X \lrcorner T)
\cdot \psi,
\label{drittelspinconn}
\ee
which is induced (as in Remark~\ref{spinorconnection_totallyanti}) by the orthogonal connection 
\bb
\widetilde{\nabla}_XY= \nabla^g_XY +3T(X,Y,\cdot)^\sharp.
\label{drittelorthconn}
\ee
We notice that the trace of $dT$ taken over the spinor space is zero due to Clifford relations.
Inserting this into the general formulas for the Seeley-deWitt coefficients (see~\cite[Theorem~4.1.6]{Gilkey95}) the claim follows.
\hfill$\boxbox$}
\medskip

\noindent
If we consider the spectral action given by the $a_2(D^2)$ and variations with respect to the torsion $3$-form $T$ we find that $T=0$ is the only possibility for critical points. 
Therefore this spectral action detects the Dirac operator induced by the Levi-Civita connection within the class of Dirac operators induced by orthogonal connections without vectorial torsion.
We note that this holds in any dimension.
This is in complete accordance with~\cite[Section~18.2]{ConnesMarcolli}.
\medskip

\noindent
The computation of $a_4(D^2)$ is more involved, we will give it only for $4$-dimensional manifolds.
In~\cite{IochumLevy} it has been noted that some terms given in~\cite{Torsion} vanish.
Similar results have been found before (compare~\cite{Goldthorpe}, \cite{Obukhov}, \cite{Grensing}).
The calculation given below is elementary, it takes place essentially in the tangent bundle and should therefore be easily accessible.

\begin{prop}
If $M$ is $4$-dimensional, the third Seeley-deWitt coefficient is 
\bb
a_4(D^2) =  \tfrac{11}{720} \, \chi(M) - \tfrac{1}{320 \pi^2} \int_M \|C\|^2 \, dvol
- \tfrac{3}{32 \pi^2} \int_M  \| \delta T \|^2  \, dvol,
\label{a4}
\ee
where $C$ is the Weyl curvature of $M$ (computed from the Levi-Civita connection).
\label{a4theorem}
\end{prop}
\pf{
We read (\ref{TorsionsBochner}) as $D^2=\Delta- E$ with potential $E = - \frac{3}{2}\, dT - \frac{1}{4}\, R^g + \frac{3}{4}\, \|T\|^2$.
From~\cite[Theorem~4.1.6,c)]{Gilkey95} we get
\[
a_4 (D^2) =   \tfrac{1}{5760 \pi^2}  \int_M  \Big(  \tr \big( 60 R^g E + 180 E^2 +
30 \,\sum_{i,j} \Omega_{ij} \Omega_{ij}\big)
 + 20\,( R^g)^2 - 8 \|\ric^g\|^2 + 8 \|\Riem^g\|^2 \Big) \, dvol\, ,
\]
where we have omitted the terms that integrate to zero over the closed manifold $M$ (Laplacians of functions).
The term $\Omega_{ij}$ is the curvature endomorphism for the spinor connection $\widetilde{\nabla}$.
The traces over the spinor space are
\begin{align*}
\tr(E)&=-R^g +3\,\|T\|^2\,,\\
\tr(E^2)&= \frac{1}{4}\,(R^g)^2-\frac{3}{2}\,R^g\,\|T\|^2+\frac{9}{4}\,\|T\|^4+\frac{9}{24}\|dT\|^2,
\end{align*}
where we note that $\tr(\omega^2)=\frac{1}{6} \,\|\omega\|^2$ for any $\omega\in\bigwedge^4$.
For the orthogonal connection $\widetilde{\nabla}$ on the tangent bundle we denote the Riemannian curvature by $\Riem$ and apply
Corollary~\ref{Omega-trace-Riem}.
Then we get
\bb
a_4(D^2) &=& \tfrac{1}{16 \pi^2}\, \tfrac{1}{360} \,
\int_M \big( 5\,(R^g)^2 - 8 \,\left\| \Ric^g \right\|^2 
+ 8 \,\left\| \Riem^g \right\|^2 
- 15\, \left\| \Riem \right\|^2   \nonumber\\
&& \qquad \qquad \quad -90\, R^g\,\|T\|^2 +405\,\|T\|^4 +\tfrac{135}{2} 
\left\|dT \right\|^2 \big) dvol
\nonumber \\
&=&  \tfrac{1}{16 \pi^2}\,\tfrac{1}{360} \, \int_M \left( 5\,(R^g)^2 - 8\, \left\| \Ric^g \right\|^2 
- 7\,\left\| \Riem^g \right\|^2 \right) dvol
\nonumber \\
&& \qquad - \tfrac{1}{16 \pi^2} \tfrac{1}{8}\, 
\int_M \left( R^g \, \left\|T \right\|^2 
-\tfrac{3}{4} \left\|dT \right\|^2  +12 \,\BB(T) +\tfrac13\, \left\|\riem^A \right\|^2 \right) dvol
\eee
by means of Proposition~\ref{theoremriem}.
With Lemma~\ref{Eulerspezial} we identify the first integral as the Euler characteristic plus the square of the Weyl curvature.
Lemma~\ref{Eulervanish} shows that the second integral equals
\[ \tfrac{1}{16 \pi^2} \tfrac{1}{8}\, 
\int_M \left( R^g \, \left\|T \right\|^2 
-\tfrac{3}{4} \left\|dT \right\|^2  +12 \,\BB(T) +\tfrac13\, \left\|\riem^A \right\|^2 \right) dvol
= \tfrac{1}{16 \pi^2} \tfrac{3}{2} \int_M  \| \delta T \|^2 \, dvol.
\]
This finishes the proof.
\hfill$\boxbox$}\medskip

\noindent
Next, we want to consider the Chamseddine-Connes spectral action (see \cite{ConnesChamseddine1}) for the Dirac operator $D$.
For $\Lambda >0$ it is defined as
\bb
I_{CC} = {\rm Tr}\, F \left( \frac{D^2}{\Lambda^2} \right) 
\nonumber
\ee
where ${\rm Tr}$ denotes the operator trace over $\HH$ as before,
and $F:\R^+\to\R^+$ is a cut-off function with support in the interval $[0,+1]$ which is constant near the origin.
Using the heat trace asymptotics one gets an asymptotic expression for $I_{CC}$ as $\Lambda\to\infty$ (see \cite{ConnesChamseddine2} for details):
\bb
I_{CC} = {\rm Tr} \, F \left( \frac{D^2}{\Lambda^2} \right) =
\Lambda^4 \, F_4 \, a_0 (D^2) + \Lambda^2 \, F_2 \, a_2(D^2)  
+ \Lambda^0 \, F_0 \, a_4(D^2) + \OO(\Lambda^{-\infty})
\label{specact}
\ee
with the  first three moments of the cut-off function which are given
by
$F_4 = \int_0^\infty s \cdot F(s) \, ds$, $F_2 = \int_0^\infty F(s) \, ds$ and $F_0 = F(0)$.
Note that these moments are independent of the geometry of the manifold.
\medskip

\noindent
Now we want to deduce the equation of motion for $I_{CC}$.
In analogy to the Riemannian Einstein-Hilbert case we consider variations with respect to the metric and the torsion $3$-form while keeping the volume fixed.
Then $a_0(D^2)$ and $\chi(M)$ are constant and their variation vanishes.
In order to avoid further complications we neglect the contributions of the $\OO(\Lambda^{-\infty})$-term.
Therefore we will consider the following action functional
\bb
{\widetilde I}_{CC}=-\alpha \int_M R^g dvol -\beta \int_M \| C\|^2 dvol+\gamma_1 \int_M \| T\|^2 dvol -\gamma_2 \int_M \| \delta T\|^2 dvol,
\label{specacttruncated}
\ee
where $\alpha,\beta,\gamma_1,\gamma_2>0$. 
\medskip

\noindent
Before we proceed let us briefly recall the standard scalar product on $k$-forms induced by a Riemannian metric $g$ (compare e.g.~with~\cite{Bleecker}).
Let $E_1,\ldots,E_n$ be an orthonormal basis of some tangent space $T_pM$.
Then the scalar product on $\bigwedge^k T^*_pM$ is uniquely determined by the requirement that $E^*_{i_1}\wedge\ldots\wedge E^*_{i_k}$, $i_1<\ldots<i_k$, form an orthonormal basis.
In local coordinates $(x^1,\ldots,x^n)$ it can be written as follows: for $\omega,\eta\in\bigwedge^k T^*_pM$ with 
\[
\omega =\sum_{i_1<\ldots<i_k}\omega_{i_1\ldots i_k} dx^{i_1}\wedge\ldots\wedge dx^{i_k},\quad
\eta =\sum_{i_1<\ldots<i_k}\eta_{i_1\ldots i_k} dx^{i_1}\wedge\ldots\wedge dx^{i_k},
\]
where the coeffients $\omega_{i_1\ldots i_k}, \eta_{i_1\ldots i_k}$ are anti-symmetric in the indices $i_1\ldots i_k$,
the scalar product is then given as
\bb\label{scalarproduct_forms}
\langle \omega,\eta\rangle_g = \frac{1}{k!} \sum_{\substack{i_1,\ldots ,i_k \\ j_1,\ldots ,j_k}} g^{i_1 j_1}\cdots g^{i_k j_k}\omega_{i_1\ldots i_k}\eta_{j_1\ldots j_k}.
\ee
Referring to the norm of $2$-forms and $3$-forms as used above we note that
$
\langle S,S\rangle_g = \tfrac12 \|S\|^2$,
$\langle T,T\rangle_g = \tfrac16 \|T\|^2$
for $S\in \bigwedge^2 T^*_pM$ and $T\in \bigwedge^3 T^*_pM$, compare (\ref{relationscalarprods}).
With respect to this scalar product the Hodge $*$-operator is an isometry, and on a $4$-manifold the $L^2$-adjoint of $d$ is $\delta=-*d*$ independently of the degree $k$ of the form. 
For $3$-forms $\delta$ is given as in (\ref{exterior_codifferential}).\medskip

\noindent
Let $g$ be a Riemannian metric on $M$.
For any $k$-form $\eta$ on $M$ we define a symmetric $(2,0)$-tensor $g^\eta$ 
for $k \geq 2$ by 
\bb
g^\eta(X,Y)=\langle X\lrcorner \eta, Y \lrcorner \eta \rangle_g 
\nonumber
\ee
and for $k=1$ by
\bb 
g^\eta(X,Y)= X\lrcorner \eta \cdot Y \lrcorner \eta = \eta(X) \, \eta(Y)  \quad \mbox{ for any tangent vectors }X,Y.
\nonumber
\ee
For $(2,0)$-tensors $a$ and $h$ the natural scalar product defined in (\ref{deftensorprod}) reads as
\bb\label{def_g-trace}
\langle a,h \rangle = \sum_{\substack{r,s\\i,j}} a_{ij} g^{ir} g^{js} h_{rs},
\ee
in local coordinates, as above.

\begin{lemma}
\label{variierealles}
Let $\left(g(t)\right)_t$ be a smooth family of Riemannian metrics on $M$ with $g(0)=g$ and ${\dot g}(0)=h$, and let $k\ge 1$.
Then for any $k$-form $\eta$ on $M$ we get
\[
\frac{d}{dt}\big|_{t=0} \langle \eta,\eta\rangle_{g(t)} = -\langle g^\eta , h\rangle.
\]
\end{lemma}
\pf{ In coordinates we write $\eta =\sum\limits_{i_1<\ldots<i_k}\eta_{i_1\ldots i_k} dx^{i_1}\wedge\ldots\wedge dx^{i_k}$ with 
$\eta_{i_1\ldots i_k}$ anti-symmetric in the indices.
We recall that $\frac{d}{dt}\big|_{t=0} g^{ij}(t)=-\sum_{r,s} g^{ir}h_{rs}g^{js}$ and use (\ref{scalarproduct_forms}) to obtain
\begin{eqnarray*}
\frac{d}{dt}\big|_{t=0} \langle \eta,\eta\rangle_{g(t)}
&=& - \frac{1}{k!} \sum_{r,s} 
\Big(
\sum_{\substack{i_1,\ldots ,i_k \\ j_1,\ldots ,j_k}} \eta_{i_1\ldots i_k} \eta_{j_1\ldots j_k}
\sum_{m=1}^k g^{i_1 j_1}\cdots {\widehat{g^{i_m j_m}}} \cdots g^{i_k j_k} g^{i_m r}g^{j_m s}
\Big)h_{rs}\\
&=& - \frac{1}{(k-1)!} \sum_{r,s} 
\Big(
\sum_{\substack{i_1,\ldots ,i_k \\ j_1,\ldots ,j_k}} \eta_{i_1\ldots i_k} \eta_{j_1\ldots j_k}
g^{i_2 j_2}\cdots g^{i_k j_k}
\Big)
g^{i_1 r}g^{j_1 s}
h_{rs}\\
&=&
-\sum_{\substack{r,s\\i_1,j_1}} (g^\eta)_{i_1 j_1} g^{i_1 r}g^{j_1 s}
h_{rs}\\
&=& -\langle g^\eta , h\rangle,
\end{eqnarray*}
where we have used the total anti-symmetry of $\eta_{i_1\ldots i_k}$ in the indices for the second equality.
{\hfill$\boxbox$}}
\begin{thm}\label{Equation_of_Motion}
Any critical point $(M,g,T)$ of $ {\widetilde I}_{CC}$ satisfies
\begin{align}
0&= 3\gamma_1\, T -\gamma_2 \,d\delta T \label{variation_torsion} \\
0&= \alpha G^g - \beta B^g 
+\gamma_1\big( -6 g^{*T}+\frac12 \|T\|^2 g\big)
-\gamma_2\big(- 2 g^{d*T}+\frac12 \|\delta T\|^2 g\big) \label{variation_metric}
\end{align}
where $G^g=\ric^g-\frac12 R^g \,g$ is the Einstein tensor of the metric $g$ and $B^g$ denotes its Bach tensor (for a definition see \cite[(4.77.)]{Besse}).
\end{thm}
\pf{
Let $(M,g,T)$ be a critial point.
We consider an arbitrary variation $T(t)$ of $3$-forms with $T(0)=T$ and ${\dot T}(0)=\tau$.
Then we have
\begin{align*}
0&= \frac{d}{dt}\big|_{t=0} \int_M \big(\gamma_1 \|T(t)\|^2 -\gamma_2 \|\delta T(t)\|^2 \big)dvol \\
&= \frac{d}{dt}\big|_{t=0} \int_M \big(6\gamma_1 \langle T(t),T(t)\rangle_g -2\gamma_2 \langle \delta T(t),\delta T(t)\rangle_g \big)dvol \\
&= \int_M \langle 3\gamma_1\, T -\gamma_2 \,d\delta T,4\tau \rangle_g dvol 
\end{align*}
since $d$ is the adjoint of $\delta$.
As $\tau$ can be chosen arbitrarily we have established (\ref{variation_torsion}).\medskip

\noindent
Now we fix $T$ and consider an arbitrary variation $g(t)$ of Riemannian metrics with $g(0)=g$ and $\dot g(0)=h$.
In the following we label any object which depends on $g(t)$.
First we note that 
$ \frac{d}{dt} |_{t=0} dvol_{g(t)} = \langle g,h\rangle dvol_g$
and by help of Lemma \ref{variierealles} we compute
\begin{align}
\frac{d}{dt}\big|_{t=0} \int_M\|T\|^2_{g(t)} dvol_{g(t)} 
&= \frac{d}{dt}\big|_{t=0} \int_M 6\left\langle T,T \right\rangle_{g(t)}
dvol_{g(t)}\nonumber \\
&= \frac{d}{dt}\big|_{t=0} \int_M 6\left\langle *_{g(t)}\,T,*_{g(t)}\,T \right\rangle_{g(t)}
dvol_{g(t)}\nonumber \\
&= 3\int_M \left(\left\langle -2 g^{*_gT}
+ \left\langle*_g T , *_g T\right\rangle_g   g,h \right\rangle
+4 \big\langle *_g T , \frac{d}{dt}\big|_{t=0} \left(*_{g(t)}\,T \right)\big\rangle_g
\right)
dvol_g\nonumber \\
&=\int_M \left(
\left\langle -6 g^{*_gT}
+ \frac12 \| T\|^2 \, g,h \right\rangle
+12 \big\langle *_g T , \frac{d}{dt}\big|_{t=0} \left(*_{g(t)}\,T \right)\big\rangle_g
\right)
dvol_g \label{dreck_1}
\end{align}
Now we calculate
\begin{align}
\frac{d}{dt}\big|_{t=0} \int_M\|\delta_{g(t)}T\|^2_{g(t)} dvol_{g(t)}
&= \frac{d}{dt}\big|_{t=0} \int_M 2\left\langle\delta_{g(t)} T,\delta_{g(t)} T \right\rangle_{g(t)}
dvol_{g(t)}\nonumber \\
&= \frac{d}{dt}\big|_{t=0} \int_M 2\left\langle d*_{g(t)} T,d*_{g(t)} T \right\rangle_{g(t)}
dvol_{g(t)}\nonumber \\
&= \int_M \Big(\left\langle -2 g^{d*_gT}
+ \left\langle d*_g T ,d *_g T\right\rangle_g   g,h \right\rangle\nonumber \\
&\qquad\qquad\quad +4 \big\langle \delta_g d*_g T , \frac{d}{dt}\big|_{t=0} \left(*_{g(t)}\,T \right)\big\rangle_g
\Big)
dvol_g\nonumber \\
&= \int_M \Big(\left\langle -2 g^{d*_gT}
+ \frac12 \| \delta_g T \|  \, g,h \right\rangle\nonumber \\
&\qquad\qquad\quad +12\,\frac{\gamma_1}{\gamma_2}\,\big\langle *_g T , \frac{d}{dt}\big|_{t=0} \left(*_{g(t)}\,T \right)\big\rangle_g
\Big)
dvol_g\label{dreck_2}
\end{align}
where we have inserted $3\gamma_1*_g\! T=\gamma_2\,\delta_g d *_g\! T$ which we obtained from (\ref{variation_torsion}).\medskip

\noindent
Finally \cite[Proposition~4.17]{Besse} and \cite[(4.77)]{Besse} tell us that
\begin{equation}
\frac{d}{dt}\big|_{t=0} \int_M \left(-\alpha R^{g(t)}-\beta \|C^{g(t)}\|^2_{g(t)}\right)
dvol_{g(t)}
=
\int_M\left\langle \alpha G^g-\beta B^g , h \right\rangle
dvol_g.\label{dreck_3}
\end{equation}
Combining (\ref{dreck_1}), (\ref{dreck_2}) and (\ref{dreck_3}) gives the assertion (\ref{variation_metric}).
{\hfill$\boxbox$}}

\begin{rem}
a) Equation (\ref{variation_torsion}) is a Proca equation for a $3$-form.
This suggests a physical interpretation of torsion as a massive vector boson.
This feature has been observed earlier in the Lorentzian context, see~\cite{Obukhov2}, and appears to be natural for dynamical Lagrangians of the torsion.\\
b) Equivalently, the Proca equation (\ref{variation_torsion}) can be expressed as
$\Delta T= \frac{\gamma_2}{3 \,\gamma_1} T $ under the condition that $dT=0$.
\end{rem}

\begin{rem}
Ricci flat manifolds $(M,g)$ with $T=0$ are critical points of ${\widetilde I}_{CC}$. 
This follows from the fact that Ricci flat manifolds have vanishing Bach tensors (see \cite[Prop.~4.78]{Besse}).
\end{rem}
Finding solutions for the equations of motions (\ref{variation_torsion}) and (\ref{variation_metric}) with $T\ne 0$ is a challenge.
The following lemmas show that  classes of warped products with special choices of $T$ can be excluded.
\begin{lemma}\label{lemma_Robertson_Walker}
Let $(N,h)$ be a compact oriented $3$-dimensional Riemannian manifold with constant curvature, and
let $f:S^1\to (0,\infty)$ be some smooth function on the circle $S^1= \R\slash \Z$.
Consider $M=S^1\times N$ equipped with the warped product metric $g=dt^2\oplus f(t)^2\,h$.
Let $\tau:M\to\R$ be a smooth function and set the $3$-form $T=\tau\cdot \pi^*dvol_{(N,h)}$ where $dvol_{(N,h)}$ is the Riemannian volume form of $(N,h)$ and $\pi:S^1\times N\to N$ is the canonical projection, and let this triple $(M,g,T)$ solve the equations of motions (\ref{variation_torsion}) and (\ref{variation_metric}). Then the torsion is zero: $T=0$.
\end{lemma}
\pf{
As $(N,h)$ is locally conformally flat, so is the warped product $(M,g)$.
Therefore the Bach tensor of $(M,g)$ vanishes, $B^g=0$.
In order to get the Einstein tensor $G^g$ we need to calculate the curvatures of $M$.
By $X$, $Y$, $Z$ we denote vectors tangent to the leaves $\{t\}\times N$, $\frac{\partial}{\partial t}$ is the unit normal vector field.
For the Levi-Connection connection we obain
\[
\nabla^g_X \tfrac{\partial}{\partial t}=\tfrac{\dot f(t)}{f(t)}\, X,\quad
\nabla^g_{\tfrac{\partial}{\partial t}} \tfrac{\partial}{\partial t} =0,\quad
\nabla^g_X Y =\nabla^h_X Y  - \dot f(t) \,f(t)\,h(X,Y)\,\tfrac{\partial}{\partial t},
\]
where $X,Y$ are also considered as tangent vectors fields of $(N,h)$ and $\nabla^h$ is the corresponding Levi-Connection connection. 
The Riemann tensor of $\nabla^g$ is given by
\begin{eqnarray*}
\Riem^g (X,Y)Z&=&\Riem^h (X,Y)Z +\left(\tfrac{\dot f(t)}{f(t)} \right)^2\cdot \left\{g(X,Z)Y- g(Y,Z)X \right\}\\
\Riem^g (X,\tfrac{\partial}{\partial t})Y&=& \tfrac{\ddot f(t)}{f(t)}\, g(X,Y)\, \tfrac{\partial}{\partial t}\\
\Riem^g (X,\tfrac{\partial}{\partial t})\tfrac{\partial}{\partial t}&=& -\tfrac{\ddot f(t)}{f(t)}\, X.
\end{eqnarray*}
From that we get the Ricci curvature and the scalar curvature 
\begin{eqnarray*}
\ric^g &=& 
\left(-3\tfrac{\ddot f(t)}{f(t)}\right)\,dt^2 \;\oplus\;\left(\ric^h -(\ddot f(t)\,f(t)+2(\dot f(t))^2)\,h \right)
\\
R^g &=&\tfrac{1}{f(t)^2}\left( R^h -6\ddot f(t)\,f(t) -6 (\dot f(t))^2  \right).
\end{eqnarray*}
As $(N,h)$ is assumed to have constant curvature there is a $\kappa\in\R$ such that
$\ric^h=2\kappa\,h$ and $R^h= 6\kappa$.
Hence the Einstein tensor of $(M,g)$ is 
\bb\label{warped_einstein}
 G^g= \tfrac{3}{f(t)^2}\left( (\dot f(t))^2-\kappa \right) dt^2\;\oplus\;
\left( 2\ddot f(t)\,f(t)+ (\dot f(t))^2-\kappa \right)\, h.
\ee
Next, we consider normal coordinates $(x^1,x^2,x^3)$ of $N$ about $p$ such that $\tfrac{\partial}{\partial x^1}, \tfrac{\partial}{\partial x^2},\tfrac{\partial}{\partial x^3}$ forms an orthonormal basis of $T_pN$ with respect to $h$.
For the volume distortion $\sqrt{h}$, given by $\sqrt{h}(x)=\sqrt{\det(h_{ij}(x)}$, we get in $p$:
\[ \sqrt{h}=1,\quad \tfrac{\partial}{\partial x^i}\sqrt{h} =0,\quad\sum_{i=1}^3\tfrac{\partial^2}{(\partial x^i)^2}\sqrt{h}=-\tfrac{1}{3} R^h(p)= -\tfrac{1}{3}\kappa.
\]
Now we take the product chart $(t,x^1,x^2,x^3)$ of $M$. 
In these coordinates the torsion $3$-form $T$ reads as $\tau(t,x^1,x^2,x^3)\sqrt{h}(x^1,x^2,x^3) dx^1\wedge dx^2\wedge dx^3$.
From (\ref{variation_torsion}) we get $dT=0$ which implies that $\tfrac{\partial \tau}{\partial t}\equiv 0$.\medskip

\noindent
As $dx^1,dx^2,dx^3$ is an orthonormal basis of $T^*_pN$ w.r.t.~$h$, we get that $dt, f\,dx^1, f\,dx^2, f\,dx^3$ forms an orthonormal basis of $T^*_{(t,p)}M$ w.r.t.~$g$.
Hence, we get
\bb\label{tosionsnormquadrat}
\| T\|_g^2=6\langle T,T\rangle = \tfrac{6\tau^2}{f^6}.
\ee
In $(t,p)$ we get $*T= \tau(x^1,x^2,x^3)\tfrac{1}{f(t)^3} dt$ and thus
\bb\label{gsternT}
g^{*T}= \tfrac{\tau^2}{f(t)^6}\,dt^2
\ee
Furthermore we obtain $d*T=\tfrac{1}{f(t)^3}\sum\limits_{i=1}^3\tfrac{\partial \tau}{\partial x^i} dx^i\wedge dt$, and therefore
\bb
g^{d*T}(\tfrac{\partial }{\partial t}, \tfrac{\partial }{\partial t})&=& \tfrac{1}{f(t)^6}\,\|\grad^g\tau\|_g^2\nonumber\\
g^{d*T}(\tfrac{\partial }{\partial t},\tfrac{\partial }{\partial x^i}) &=& 0\nonumber\\
g^{d*T}(\tfrac{\partial }{\partial x^i},\tfrac{\partial }{\partial x^j})&=& \tfrac{1}{f(t)^6}\, d\tau(\tfrac{\partial }{\partial x^i})\cdot  d\tau(\tfrac{\partial }{\partial x^j}).
\eee
If we now assume that equation (\ref{variation_metric}) holds, we observe that after restricting the occuring $(2,0)$-tensors to $TN$ every ingredient is a multiple of $h$ except $g^{d*T}$. So (\ref{variation_metric}) can only hold if $d\tau=0$.
Therefore the function $\tau$ is constant, and $d*T=0$ and so $g^{d*T}=0$.
Then we decompose (\ref{variation_metric}) into its $dt^2$-component and its $h$-component:
\bb
 \tfrac{3\alpha}{f(t)^2}\left( \dot f(t))^2-\kappa\right)&=& \tfrac{6\gamma_1\tau^2}{f(t)^6}- \tfrac{6\gamma_1\tau^2}{2f(t)^6}\nonumber\\
 \alpha\left( 2\ddot f(t)\,f(t)+ (\dot f(t))^2-\kappa \right)&=&- \tfrac{6\gamma_1\tau^2}{2f(t)^4} 
\eee
We divide the first equation by $3$, the second one by $(f(t)^2$, and substract the first from the second.
We get:
\[
\ddot f(t) = - \tfrac{\gamma_1\tau^2}{\alpha}\cdot \tfrac{1}{f(t)^5}.
\]
Since $f>0$  and $\alpha > 0$ we conclude that $\ddot f\geq 0$ and therefore $\dot f$ increases monotonically.
On the other hand $\dot f$ being a function on $S^1=\R\slash\Z$ is periodic.
Therefore $\dot f$ is constant and hence $\tau$ is zero.
{\hfill$\boxbox$}}
\begin{rem}
It should be noted that with the ansatz in Lemma~\ref{lemma_Robertson_Walker}  one can obtain 
restrictions on the geometry of $(N,h)$ pointwise from equation (\ref{variation_torsion}). Namely, in the normal coordinates from above one finds
\[
 d\delta T = -\tau \sum_{i=1}^3 \tfrac{\partial^2}{(\partial x^i)^2}\sqrt{h}=\tfrac{1}{3} \kappa T.
\]
By (\ref{variation_torsion}) we have $\tfrac{1}{3} \kappa = \tfrac {3\gamma_1}{\gamma_2}>0$, therefore $(N,h)$ is a spherical space form.
\end{rem}
\begin{rem}
If we now consider formally the same equations of motions for Lorentzian manifolds, and if we admit also non-compact manifolds as solutions, the argument from 
the proof of Lemma~\ref{lemma_Robertson_Walker} cannot discard the Robertson-Walker ansatz $M=\R\times N$ with $g= -dt^2\oplus f(t)^2\,h$ and 
$T=\tau\cdot \pi^*dvol_{(N,h)}$ because
in the above proof the compactness of $S^1$ is essential.
\end{rem}
One could argue that in the above examples the torsion $T$ is not dynamical.
In the last example we consider a situation where torsion may be dynamical, and we show that the torsion vanishes by other reasons.
\begin{lemma}\label{warpedtorsiontorus}
Let $(N,h)$ be the flat torus $T^3=\R^3\slash\Z^3$, let $f:S^1\to (0,\infty)$ be some smooth function.
Consider $M=S^1\times N$ equipped with the warped product metric $g=dt^2\oplus f(t)^2\,h$.
Let $\tau:S^1\to (0,\infty)$ be a smooth function, and let 
\[ 
T=\tau(t) (dx^1\wedge dx^2 +dx^2\wedge dx^3 + dx^3\wedge dx^1)\wedge dt.
\]
Assume that the triple $(M,g,T)$ solves the equations of motions (\ref{variation_torsion}) and (\ref{variation_metric}). 
Then the torsion is zero: $T=0$.
\end{lemma}
\pf{
For the $T$ as above we find
\bb
*T&=& \tfrac{\tau(t)}{f(t)}\left(dx^3+dx^1+dx^2\right), \nonumber\\
d*T&=&\tfrac{f(t)\dot\tau(t)-\tau(t)\dot f(t)}{f(t)^2} dt\wedge \left(dx^3+dx^1+dx^2\right).
\eee
Furthermore we get $\| T\|^2=18\cdot (\tfrac{\tau(t)}{f(t)^2})^2$ and $\|\delta T\|^2= \tfrac{2}{f(t)^6}\cdot\left(f(t)\dot\tau(t)-\dot f(t)\tau(t)\right)^2$ and
\bb
g^{*T}(\tfrac{\partial}{\partial t},\tfrac{\partial}{\partial t})&=& 0,\nonumber\\
g^{d*T}(\tfrac{\partial}{\partial t},\tfrac{\partial}{\partial t})&=& 3\cdot \tfrac{(f(t)\dot\tau(t)-\tau(t)\dot f(t))^2}{f(t)^6}.
\eee
Now we insert twice $\tfrac{\partial}{\partial t}$ into the $(2,0)$-tensors of (\ref{variation_metric}) and use the specific form of the Einstein tensor (\ref{warped_einstein}). This yields 
\[
0= 3\alpha \cdot \tfrac{\left( \dot f(t)\right)^2}{\left(f(t) \right)^2} + 9\gamma_1 \cdot \tfrac{\left(\tau(t)\right)^2}{\left(f(t)\right)^4} 
+ 5\gamma_2 \cdot  \tfrac{\left(f(t) \dot \tau(t) - \tau(t) \dot f(t)\right)^2}{\left(f(t)\right)^6}.
\]
Since $\alpha,\gamma_1,\gamma_2 > 0$ each summand is nonnegative, therefore each term vanishes individually, in particular the second one.
From this  we conclude $\tau=0$.
{\hfill$\boxbox$}}
\begin{rem}
In the Lorentzian setting actions similar to $\widetilde{I}_{CC}$ have already been consider and some cosmological consequences for possible critical points with non-vanishing torsion have been discussed (see e.g.~\cite{HHKN76} or \cite{S02} and the references therein).
\end{rem}


\appendix

\section{Appendix: Proofs of Proposition \ref{theoremriem},  Theorem \ref{Eulerallg} 
and Lemma \ref{intzero}}

We consider a manifold $M$ equipped with a Riemannian metric $g$.
For $(k,0)$-tensor fields $S$, $T$ on $M$ one has pointwise the natural scalar product
\bb
\langle S,T \rangle = \sum_{i_1,\ldots,i_k} S(E_{i_1},\ldots, E_{i_k}) \cdot T(E_{i_1},\ldots, E_{i_k}),
\label{deftensorprod}
\ee
where $(E_i)_i$ is an orthonormal basis of the tangent space. This coincides with the definition given in 
(\ref{30tensorprod}) for $(3,0)$-tensors.
If $S$ and $T$ are $k$-forms there is another natural scalar product, as defined in (\ref{scalarproduct_forms}),
given by
\bb
\langle S,T \rangle_g &=& \sum_{i_1<\ldots<i_k} S(E_{i_1},\ldots, E_{i_k}) \cdot T(E_{i_1},\ldots, E_{i_k})
\nonumber \\ 
&=& \frac{1}{k!}  
\sum_{i_1,\ldots,i_k} S(E_{i_1},\ldots, E_{i_k}) \cdot T(E_{i_1},\ldots, E_{i_k}) \quad =  \quad \frac{1}{k!} \langle S,T \rangle. 
\label{relationscalarprods}
\ee

\noindent
Now let $\nabla^{g}$ denote the Levi-Civita connection on the tangent bundle.
We fix some $3$-form $T$ on $M$ and some $s\in\R$, and we are studying the connection $\nabla$ with totally anti-symmetric 
torsion as in (\ref{vectortorsionconnection}).

\noindent
As in section~\ref{section1} we fix some point $p\in M$, and we extend any tangent vectors $X,Y,Z,W\in T_p M$ to vector fields again denoted by $X,Y,Z,W$ being synchronous in $p$.
\medskip

\noindent
We recall the Ricci decomposition for $\Riem^g$ (compare \cite[Chapter 1.G]{Besse}): 
Let the tracefree part of the Ricci curvature be denoted by
$
\B(X,Y)=\ric^g(X,Y)-\tfrac{1}{n}\;R^g\,\left\langle X,Y \right\rangle.
$
Then, one has
\begin{align}
\left\langle \Riem^g(X,Y)Z,W \right\rangle =& \tfrac{1}{n(n-1)}\;R^g\,
\left(\left\langle X,W \right\rangle \left\langle Y,Z \right\rangle - \left\langle Y,W \right\rangle \left\langle X,Z \right\rangle\right)\nonumber \\
&+\tfrac{1}{n-2} \big( \B(X,W)\,\left\langle Y,Z \right\rangle -\B(Y,W)\, \left\langle X,Z \right\rangle
 + \left\langle X,W\right\rangle\,\B(Y,Z)- \left\langle Y,W \right\rangle\,\B(X,Z)\big)\nonumber \\
&+ \left\langle C(X,Y)Z,W \right\rangle,\label{Ricci_decomposition}
\end{align}
where the $(3,1)$-tensor $C$ is the Weyl tensor of the Riemannian metric $g$.
The Weyl tensor possesses all the symmetries of the Riemann tensor $\Riem^g$ (e.g.~the Bianchi identity holds), and in addition the contraction of $\left\langle C(X,Y)Z,W \right\rangle$ taken over any two slots is zero.
Furthermore, it should be noted that this composition into scalar curvature, trace-free Ricci curvature and Weyl curvature is orthogonal with respect to the usual scalar product in the space of $(4,0)$-tensors (see \cite[Thm.~1.114]{Besse}).
\medskip

\noindent

\subsection{Proof of Proposition \ref{theoremriem}}

\paragraph{Proof of Proposition \ref{theoremriem}. }
First, consider some Riemannian manifold $(M,g)$ of arbitrary dimension $n$
with Levi-Civita connection $\nabla^{g}$.
We fix some $3$-form $T$ on $M$, and we are studying the connection $\nabla$ 
defined as in (\ref{vectortorsionconnection}).\medskip

\noindent
The symmetric part of $\riem$ is 
\begin{align}
\riem^S(X,Y,Z,W)&=\tfrac{1}{2}\left(\riem(X,Y,Z,W)+ \riem(Z,W,X,Y)\right)\nonumber\\
&=\riem^g(X,Y,Z,W)  + \tfrac{s}{2}\; dT(X,Y,Z,W)\nonumber\\
&\qquad+ s^2\,\left(\left\langle T(X,Z,\cdot)^\sharp,T(Y,W,\cdot)^\sharp \right\rangle - \left\langle T(Y,Z,\cdot)^\sharp,T(X,W,\cdot)^\sharp\right\rangle\right),\label{symm_Riemann}
\end{align}
where we have used (\ref{Riemann_identity}) and (\ref{exterior_differential}).
\medskip

\noindent
In the following we will abbreviate
\begin{align}
C_{ijkl}&=\left\langle C(E_i,E_j)E_k,E_l \right\rangle, \qquad
\riem^{(g/S/A)}_{ijkl}= \riem^{(g/S/A)}(E_i,E_j,E_k,E_l), \qquad
\nonumber \\
\ric^{(g/S/A)}_{ij}&=\ric^{(g/S/A)}(E_i,E_j), \qquad \B_{ij}=\B(E_i,E_j), 
\nonumber \\
T_{ij}&=T(E_i,E_j,\cdot)^\sharp , \qquad
dT_{ijkl}=dT(E_i,E_j,E_k,E_l )  
\nonumber
\end{align}
for all indices $i,j,k,l$.
\medskip

\noindent
We will calculate $\| \riem^S \|^2 = \sum_{i,j,k,l=1}^n \left(\riem^S_{ijkl}\right)^2$ by means of (\ref{symm_Riemann}).
Three cross terms appear, one of them vanishes in the case of arbitrary dimension $n$:
We choose some orthonormal frame $E_1,\ldots,E_n$, which is defined on some neighbourhood of $p$ and is synchronous in $p$. 
Using the Bianchi identity we compute
\begin{equation}\label{crossterm1}
0=\tfrac{1}{3}\sum_{i,j,k,l=1}^n \left(\riem^g_{ijkl}+\riem^g_{jkil}+\riem^g_{kijl} \right)\, dT_{ijkl}
= \sum_{i,j,k,l=1}^n \left(\riem^g_{ijkl}\right)\, dT_{ijkl}.
\end{equation}
\noindent
From now on we restrict to the case of dimension $n=4$.
We note that $dT_{ijkl}\ne 0$ is only possible if $i,j,k,l\in\{1,\ldots,4\}$ are pairwise distinct. 
In that case one has 
$
\left\langle T_{ij}, T_{kl} \right\rangle=\sum_{m=1}^4 T(E_i,E_j,E_m)\, T(E_k,E_l,E_m)=0
$
as any $m\in\{1,\ldots,4\}$ is equal to one of $i,j,k,l$.
Hence we get
\begin{equation}
0= \sum_{i,j,k,l=1}^4 dT_{ijkl}\cdot
 \Big( 
\left\langle T_{ik}, T_{jl} \right\rangle- 
\left\langle T_{jk}, T_{il}\right\rangle
\Big).
\label{crossterm2}
\end{equation}
In order to identify the third cross term we impose the Ricci decomposition (\ref{Ricci_decomposition}) which now reads as
\bb
\riem^g_{ijkl}=
\tfrac{1}{12}\,R^g\,\left(\delta_{ij}\delta_{jk}-\delta_{jl}\delta_{ik} \right)
+ \tfrac{1}{2}\,\big( \B_{il}\,\delta_{jk}-\B_{jl}\,\delta_{ik} +
\delta_{il}\,\B_{jk}-\delta_{jl}\,\B_{ik}\big) 
+ C_{ijkl}
\eee
and we get
\bb
\sum_{i,j,k,l}\left(\riem^g_{ijkl}\right)\,
\left(
\left\langle T_{ik},T_{jl} \right\rangle -
\left\langle T_{jk},T_{il} \right\rangle
\right)
&=&  \tfrac{1}{12}\,R^g\,\sum_{i,j}
\left(
\left\langle T_{ij},T_{ji} \right\rangle +
\left\langle T_{ji},T_{ij} \right\rangle
\right)
\nonumber \\
&& + \tfrac{1}{2}\,\Big(
\sum_{i,j,l} \B_{il}\,\left\langle T_{ij},T_{jl} \right\rangle
+
\sum_{i,j,,l}\B_{jl}\,\left\langle T_{ji},T_{il} \right\rangle
\nonumber \\
&& \qquad+ 
\sum_{i,j,k}\B_{jk} \,\left\langle T_{ik},T_{ji} \right\rangle
+
\sum_{i,j,k}\B_{ik}\,\left\langle T_{jk},T_{ij} \right\rangle
\Big)
\nonumber \\
&& +\sum_{i,j,k,l} C_{ijkl} \, 
\left(
\left\langle T_{ik},T_{jl} \right\rangle -
\left\langle T_{jk},T_{il} \right\rangle
\right)
\nonumber \\
&=&   -\tfrac{1}{6}\,R^g\,\left\|T\right\|^2
+2\,\BB(T) +\CC(T)
\eee
where we have set
\bb
\BB(T)&=& \sum_{i,j,k} \B_{ik}\, 
\left\langle T_{ij},T_{jk} \right\rangle 
= \sum_{i,j,k} \ric^g_{ik}\, 
\left\langle T_{ij},T_{jk} \right\rangle
-\tfrac{1}{4}\,R^g\,\sum_{i,j}\left\langle T_{ij},T_{ji} \right\rangle\nonumber \\
&=& \sum_{i,j,k} \ric^g_{ik}\, 
\left\langle T_{ij},T_{jk} \right\rangle
+\tfrac{1}{4}\,R^g\,\left\| T\right\|^2\label{B_von_T} \\
\CC(T)&=& \sum_{i,j,k,l} C_{ijkl}\, 
\left(
\left\langle T_{ik},T_{jl} \right\rangle -
\left\langle T_{jk},T_{il} \right\rangle
\right)\label{C_von_T}
\ee
It follows that
\bb
\| \riem^S \|^2 &=& \| \riem^g\|^2 + \tfrac{1}{4} s^2 \| dT \|^2
-\tfrac{1}{3}s^2 \,R^g\,\left\|T\right\|^2 +4 s^2 \,\BB(T) +s^2 \CC(T)
\nonumber \\
&&+ �s^4\,
\sum_{i,j,k,l=1}^4
\left(\left\langle T_{ik},T_{jl} \right\rangle 
- \left\langle T_{jk},T_{il} \right\rangle\right)^2
\nonumber \\
&=& \left\|\riem^g \right\|^2 
+ \tfrac{1}{4} s^2 \,  \left\|dT \right\|^2 
- \tfrac{1}{3}s^2 \,R^g\,\left\|T\right\|^2
+4 s^2 \,\BB(T) +  \tfrac{1}{3} s^4\,  \left\| T \right\|^4 
\eee
where for the last step we use the following technical Lemmas \ref{cct0} and \ref{Riemtttt}.
\hfill$\boxbox$
\medskip

\begin{lemma}
Let $M$ be a $4$-dimensional Riemannian manifold and $p\in M$.
For any $T\in\bigwedge^3T^*_pM$ we consider $\CC(T)$ as defined (\ref{C_von_T}).
Then one has $\CC(T)=0$.
\label{cct0}
\end{lemma}
\pf{
Let $E_1,E_2,E_3,E_4$ be an orthogonal basis of $T_pM$.
For $T,S\in \bigwedge^3T^*_pM$ we define
\[
 \cc(T,S)=2\cdot\sum_{i,j,k,l=1}^4 C_{ijkl}\left\langle T_{ik},S_{jl} \right\rangle.
\]
The Weyl tensor satisfies $C_{ijkl}=C_{jilk}$, and hence $\cc$ is a symmetric bilinear form.
For $T\in\bigwedge^3T^*_pM$ we have
\begin{equation*}
\cc(T,T)=\sum_{i,j,k,l} C_{ijkl} \, \left\langle T_{ik},T_{jl} \right\rangle
- \sum_{i,j,k,l} C_{jikl} \, \left\langle T_{ik},T_{jl} \right\rangle
=\sum_{i,j,k,l} C_{jikl}\,\left( \left\langle T_{ik},T_{jl}\right\rangle 
- \left\langle T_{jk},T_{il}\right\rangle \right)
=\CC(T).
\end{equation*}
Hence, the Lemma will be proved if we show that $\cc\equiv 0$.
We will check this on the basis of $\bigwedge^3T^*_pM$ consisting of
${E_1}^\flat\wedge {E_2}^\flat\wedge {E_3}^\flat$, 
${E_2}^\flat\wedge {E_3}^\flat\wedge {E_4}^\flat$, 
${E_1}^\flat\wedge {E_3}^\flat\wedge {E_4}^\flat$, 
${E_1}^\flat\wedge {E_2}^\flat\wedge {E_4}^\flat$.\medskip

\noindent
For $t_1,t_2\in \R$ we consider $T_1,T_2\in \bigwedge^3T^*_pM$ defined by
\[
T_1 := t_1 \cdot {E_1}^\flat\wedge {E_2}^\flat\wedge {E_3}^\flat, \quad
T_2 := t_2 \cdot {E_2}^\flat\wedge {E_3}^\flat\wedge {E_4}^\flat.
\]
We will verify that $\cc(T_1,T_1)=0$ and $\cc(T_1,T_2)=0$, this will suffice because the computations are the same if one inserts any elements of the above basis into $\cc$.
For $a\in \{1,2\}$ Table~\ref{tab1} gives $(T_a)_{ij} := T_a(E_i,E_j, \cdot)^\sharp$ explicitly. \medskip

\begin{table}[h]
$(T_1)_{ij}: \;$
\begin{tabular}{c|cccc}
\backslashbox[0pt][l]{i}{j} &1&2&3&4 \\
\hline
1 & 0 & $-t_1E_3$ & $t_1E_2$  & 0  \\
2 &$t_1E_3$& 0 &$-t_1E_1$ & 0 \\
3 &$-t_1E_2$ &$t_1E_1$ &0&0 \\
4 &0&0&0&0 \\ 
\end{tabular}
$\qquad\qquad (T_2)_{ij}: \;$
\begin{tabular}{c|cccc}
\backslashbox[0pt][l]{i}{j} &1&2&3&4 \\
\hline
1 & 0 & 0 & 0  & 0  \\
2 &0& 0 &$-t_2E_4$ & $t_2E_3$ \\
3 &0 &$t_2E_4$ &0&$-t_2E_2$ \\
4 &0&$-t_2E_3$&$t_2E_2$ &0 \\
\end{tabular}
\caption{$(T_a)_{ij} := T_a(E_i,E_j, \cdot)^\sharp$ for $a\in \{1,2\}$.}
\label{tab1}
\end{table}
\noindent
Using Table~\ref{tab1} we compute
\begin{equation}\label{ww0}
\cc(T_1,T_1)=\sum_{i,j,k,l=1}^4 C_{ijkl} \, \left\langle T_{ik},T_{jl} \right\rangle 
= \sum_{i,j,k,l=1}^3 C_{ijkl}\; t_1^2 \left( \delta_{ij} \, \delta_{kl} -\delta_{il} \, \delta_{jk}\right)
= -  t_1^2  \sum_{i,j=1}^3  C_{ijji}
\end{equation}
The Weyl tensor is trace-free in any pair of indices.
Hence for any $i \in \{1,2,3 \}$ we get $- C_{i44i}=\sum_{j=1}^3  C_{ijji}$.
Inserting this into (\ref{ww0}) and using $C_{4444}=0$ we obtain
\begin{equation*}
\cc(T_1,T_1)=t_1^2 \, \sum_{i=1}^4  C_{i44i}=0
\end{equation*}
as the Weyl tensor is trace-free in any pair of indices.\medskip

\noindent
With table \ref{tab1} we obtain
\bb
\cc(T_1,T_2) 
&=&  2 \sum_{i,j,k,l=1}^4 C_{ijkl} \, 
\left\langle (T_1)_{ik},(T_2)_{jl}  \right\rangle
\nonumber \\
&=& -2 \, t_1 \, t_2 \,(C_{1224} + C_{2412} + C_{1334} +C_{3413} )
\nonumber \\
&=& -4  \, t_1 \, t_2 \, ( C_{1224} + C_{1334} ) 
\nonumber \\
&=& -4  \, t_1 \, t_2 \, \sum_{i=1}^4 C_{1ii4} 
\nonumber \\
&=& 0
\eee
where we used the symmetry $C_{ijkl} = C_{klij}$ of the Weyl tensor in 
the second step and $C_{1114}=C_{1444}=0$ in the third step,
the  last step is achieved as the Weyl tensor is trace-free.
\hfill$\boxbox$
}
\medskip

\begin{lemma}
\label{Riemtttt}
Let $M$ be a Riemannian manifold of dimension $4$, let $p\in M$ and let $E_1,E_2,E_3,E_4$ be an orthonormal basis of $T_pM$.
Then for any $T\in \bigwedge^3T^*_pM$ one has 
\[
\sum_{i,j,k,l=1}^4 
\left(\left\langle T_{ik},T_{jl} \right\rangle - \left\langle T_{jk},T_{il}\right\rangle\right)^2
= \frac{1}{3} \left\| T \right\|^4.
\]
\end{lemma}
\pf{
Choosing the parameters properly we have 
$T  = T_1 + T_2 + T_3 + T_4$
with  $T_1$ and $T_2$ as in  Lemma \ref{cct0} and 
\[
T_3 := t_3 \cdot {E_1}^\flat\wedge {E_3}^\flat\wedge {E_4}^\flat, \quad 
T_4 := t_4 \cdot {E_1}^\flat\wedge {E_2}^\flat\wedge {E_4}^\flat.
\]
Defining $(T_a)_{ij}$ as in Lemma \ref{cct0} we summarise $(T_3)_{ij}$ and $(T_4)_{ij}$ in Table~\ref{tab2}.

\begin{table}[h]
$(T_3)_{ij}: \;$
\begin{tabular}{c|cccc}
\backslashbox[0pt][l]{i}{j} &1&2&3&4 \\
\hline
1 & 0 & $0$ & $-t_3E_4$  & $t_3E_3$  \\
2 &0& 0 &0 & 0 \\
3 &$t_3E_4$ &0 &0&$-t_3E_1$ \\
4 &$-t_3E_3$&0&$t_3E_1$&0 \\
\end{tabular}
$\qquad \qquad (T_4)_{ij}: \;$
\begin{tabular}{c|cccc}
\backslashbox[0pt][l]{i}{j} &1&2&3&4 \\
\hline
1 & 0 & $-t_4E_4$ & 0  & $t_4E_2$  \\
2 &$t_4E_4$ & 0 &$0$ & $-t_4E_1$ \\
3 &0 &0 &0&0  \\
4 &$-t_4E_2$ &$t_4E_1$&0 &0 \\
\end{tabular}
\caption{$(T_a)_{ij} := T_a(E_i,E_j, \cdot)^\sharp$ for $a\in \{3,4\}$.}
\label{tab2}
\end{table}
 
\noindent
In order to calculate $\| T \|^4 = \left(\| T \|^2\right)^2$ we note that
\[
\sum_{i,j=1}^4 \left\langle (T_a)_{ij}, (T_b)_{ij} \right\rangle = 
\left\{ 
\begin{array}{l}
0 \quad  {\rm if} \quad a \neq b \\ \\
6 \, t_a^2  \quad {\rm if} \quad a=b
\end{array}
\right. 
\]
and therefore 
\bb
 \left\| T \right\|^2 = \sum_{i,j=1}^4 \left\langle T_{ij}, T_{ij} \right\rangle = 
\sum_{i,j,a=1}^4 \left\langle (T_a)_{ij}, (T_a)_{ij} \right\rangle 
= 6 \sum_{a=1}^4 t_a^2.
\label{T4}
\ee
Next, we also want to express $ \sum\limits_{i,j,k,l=1}^4 
\left(\left\langle T_{ik},T_{jl} \right\rangle - \left\langle T_{jk},T_{il}\right\rangle\right)^2$ in terms of $t_1,\ldots, t_4$.\medskip

\noindent
We note that each summand $\left(\langle T_{ik},T_{jl}\rangle - \langle T_{jk},T_{il}\rangle\right)^2 $ remains the same after interchanging $i$ and $j$ and is zero if $i=j$.
Noticing the same for the indices $k$ and $l$, we restrict ourselves to the case $i<j$, $k<l$.\medskip

\noindent
For $(i,j)=(1,2)$ and $(k,l)=(1,2)$ we get
\[
\left( \left\langle T_{11}, T_{22} \right\rangle -\left\langle T_{21}, T_{12} \right\rangle \right)^2
=  \left(\left\langle T_{12}, T_{12} \right\rangle \right)^2 
=  \left( \left\langle (T_1)_{12}, (T_1)_{12} \right\rangle 
+ \left\langle (T_4)_{12}, (T_4)_{12} \right\rangle  \right)^2
= ( t_1^2 + t_4^2 )^2,
\]
where we have used that
\[
\left\langle (T_a)_{ij}, (T_b)_{ij} \right\rangle = 
\left\{ 
\begin{array}{l}
0 \quad  {\rm if} \quad a \neq b \\ \\
t_a^2  \quad {\rm if} \quad a=b.
\end{array}
\right. 
\]
For $(i,j) = (1,2)$ and $(k,l)=(1,3)$ we find
\[
\left( \left\langle T_{11}, T_{23} \right\rangle -\left\langle T_{21}, T_{13} \right\rangle \right)^2
=  \left(\left\langle T_{12}, T_{13} \right\rangle \right)^2 
\nonumber \\
=\left( \left\langle (T_4)_{12}, (T_3)_{13} \right\rangle  \right)^2
\nonumber \\
= \, t_3^2 t_4^2,
\]
and for $(i,j) = (1,2)$ and $(k,l) = (3,4)$  we have
\[
\left( \left\langle T_{13}, T_{24} \right\rangle -\left\langle T_{23}, T_{14} \right\rangle \right)^2
= 0.
\]
The same considerations apply to the other summands, and we give the results for the cases $i<j$, $k<l$ in Table~\ref{tab3}.
\begin{table}[h]
\begin{tabular}{c|cccccc}
\backslashbox[0pt][l]{$(i,j)$}{$(k,l)$} & $(1,2)$ & $(1,3)$ & $(1,4)$ 
& $(2,3)$ & $(2,4)$ & $(3,4)$ \\
\hline \\
$(1,2)$ 
& $(t_1^2 + t_4^2)^2$ &  $t_3^2 t_4^2$  &  $t_1^2 t_3^2 $ &
 $t_2^2 t_4^2 $&  $t_1^2 t_2^2$& 0 \\ \\
$(1,3)$ &  $t_3^2 t_4^2 $  & $ (t_1^2 + t_3^2)^2$ &  $ t_1^2 t_4^2$ 
& $t_2^2 t_3^2 $ & 0 &  $t_1^2 t_2^2$ \\ \\
$(1,4)$ &  $t_1^2 t_3^2$ &  $t_1^2 t_4^2$ & $ (t_3^2 + t_4^2)^2$ & 0 
&  $t_2^2 t_3^2$  &   $t_2^2 t_4^2$ \\ \\
$(2,3)$ &  $t_2^2 t_4^2$ &  $t_2^2 t_3^2$  & 0 & $ (t_1^2 + t_2^2)^2$ 
&  $t_1^2 t_4^2$ &  $t_1^2 t_3^2$   \\ \\
$(2,4)$ & $t_1^2 t_2^2$ & 0 &  $t_2^2 t_3^2$ &  $t_1^2 t_4^2$ &
$ (t_2^2 + t_4^2)^2$ &  $t_3^2 t_4^2$ \\ \\
$(3,4)$ &0&  $t_1^2 t_2^2$  &  $t_2^2 t_4^2$ &  $t_1^2 t_3^2$ &
 $t_3^2 t_4^2$ &  $ (t_2^2 + t_3^2)^2$  \\ 
\end{tabular}
\caption{The summands $\left(\left\langle T_{ik},T_{jl} \right\rangle - \left\langle T_{jk},T_{il}\right\rangle\right)^2$ for $i<j$, $k<l$.}
\label{tab3}
\end{table}

\noindent
Now we drop the condition $i<j$, $k<l$, and we get a factor $4$ when we add up all summands:
\bb
\sum_{i,j,k,l=1}^4 
\left(\left\langle T_{ik},T_{jl} \right\rangle - \left\langle T_{jk},T_{il}\right\rangle\right)^2
&=& 4 \sum_{a\neq b} (t_a^2 + t_b^2 )^2 + 16 \sum_{a\neq b} t_a^2 t_b^2
\nonumber \\
&=& 12 \sum_{a=1}^4 t_a^4 + 8 \sum_{a\neq b} t_a^2 t_b^2 
+ 16 \sum_{a\neq b} t_a^2 t_b^2
\nonumber \\
&=& 12 \big(  \sum_{a=1}^4 t_a^4 + 2 \sum_{a\neq b} t_a^2 t_b^2 \big)
\nonumber \\
&=& 12 \big( \sum_{a=1}^4 t_a^2  \big)^2,
\nonumber
\ee
from which in combination with (\ref{T4}) the claim follows.
\hfill$\boxbox$
}

\begin{lemma}
\label{riemAsquare}
The square of the anti-symmetric part of the Riemann tensor of $\nabla$
is pointwise given by 
\[
 \left\|\riem^A \right\|^2 = s^2 \, \sum_{i,j,k,l=1}^n 
\left( \left(\nabla^g_{E_i} T\right)(E_j,E_k,E_l) 
\right)^2
+ s^2\, \sum_{i,j,k,l=1}^n 
\left( \nabla^g_{E_i} T \right)(E_j,E_k,E_l) 
\, \left( \nabla^g_{E_j} T \right) (E_i,E_k,E_l)  
\]
\end{lemma}
\pf{ 
The anti-symmetric part of $\riem$ is then given by
\bb
\riem^A(X,Y,Z,W)&=&\tfrac{1}{2}\left(\riem(X,Y,Z,W)- \riem(Z,W,X,Y)\right)
\nonumber \\
&=&\tfrac{s}{2}  \; \Big(\left(\nabla^g_X T\right)(Y,Z,W) -\left(\nabla^g_Y T\right)(X,Z,W)
\nonumber \\ 
&&\qquad -\left(\nabla^g_Z T\right)(X,Y,W)+\left(\nabla^g_W T\right)(X,Y,Z)\Big).
\ee
\noindent
The lemma follows by direct calculation.
\hfill$\boxbox$
}

\noindent
This shows in particular that $\riem^A$ does only depend on the torsion $T$ and not on
the Riemannian curvature $\Riem^g$. 

\subsection{The square of the norm of the Ricci curvature}
Now we decompose the Ricci curvature 
into its symmetric and its anti-symmetric components 
\bb
\ric(X,Y) = \ric^S(X,Y) + \ric^A(X,Y).
\eee
With (\ref{Ricci_identity}) we have
\bb
\ric^S(X,Y)&=&\tfrac{1}{2}\,\left(\ric(X,Y)+\ric(Y,X) \right)
= \ric^g(X,Y)-s^2 \;\sum_{i=1}^n \left\langle T(X,E_i,\cdot)^\sharp, T(Y,E_i,\cdot)^\sharp\right\rangle
\label{symm_Ricci}
\ee
and
\bb
\ric^A(X,Y)&=&\tfrac{1}{2}\,\left(\ric(X,Y)-\ric(Y,X) \right)
=s \;\sum_{i=1}^n \left(\nabla^g_{E_i} T\right)(X,Y,E_i)
= - s\;\delta T(X,Y).
\label{antisymm_Ricci}
\ee
\noindent

\noindent
To evaluate $\| \ric \|^2$ of the Ricci curvature (\ref{Ricci_identity}) we proceed 
as in the evaluation of $\| \Riem \|^2$  and get
\bb
\| \ric \|^2 &=& \sum_{i,j=1}^n \left( \ric(E_i,E_j) \right)^2 = \sum_{i,j=1}^n 
\ric_{ij}^2
= \sum_{i,j=1}^n \left(\ric^S_{i,j}\right)^2 + \sum_{i,j=1}^n \left(\ric^A_{ij}\right)^2
\nonumber \\
&=&  \| \ric^S  \|^2 + \| \ric^A  \|^2.
\ee
We find
\begin{thm}
Let $M$ be a $4$-dimensional manifold with Riemannian metric $g$ and 
connection $\nabla$ as given in (\ref{vectortorsionconnection}).
Then the norm of the Ricci tensor of $\nabla$ is given by
\[
\left\| \ric \right\|^2 = \left\|\ric^g \right\|^2 +  \tfrac{1}{3} s^4\, \left\| T \right\|^4  
- \tfrac{1}{2} s^2 \,R^g\,\left\|T\right\|^2
+2s^2 \,\BB(T)  + s^2 \| \delta T \|^2 
\]
with $\BB(T)$ as defined in (\ref{B_von_T}).
\label{theoremric}
\end{thm}
\pf{ 
In order to compute the squared norm $\| \ric \|^2$ of the Ricci curvature defined in (\ref{Ricci_identity}) we proceed 
as in the evaluation of $\| \Riem \|^2$. 
With (\ref{symm_Ricci})  and (\ref{antisymm_Ricci}) we get
\[
\| \ric \|^2 = \sum_{i,j=1}^n \left( \ric(E_i,E_j) \right)^2 = \sum_{i,j=1}^n \ric_{ij}^2
= \sum_{i,j=1}^n \left(\ric^S_{ij}\right)^2 + \sum_{i,j=1}^n \left(\ric^A_{ij}\right)^2.
\]
We restrict ourselves again to dimension $n=4$ and calculate 
$ \| \ric^S\|^2=\sum_{i,j=1}^n \left(\ric^S_{ij}\right)^2$ with  (\ref{symm_Ricci}). We find 
\bb
 \sum_{i,j=1}^4 \left(\ric^S_{ij}\right)^2 &=&  \sum_{i,j=1}^4 \left(\ric^g_{ij}\right)^2
- 2 s^2 \, \sum_{i,j,k=1}^4 \ric^g_{ij} \, \left\langle T_{ik}, T_{jk} \right\rangle
+ s^4 \;
\sum_{i,j,k,l=1}^4 \left\langle T_{ik}, T_{jk}\right\rangle
\left\langle T_{il}, T_{jl}\right\rangle
\nonumber \\
&=& \ \left\|\ric^g \right\|^2   
- \tfrac{1}{2}s^2\,R^g\,\left\|T\right\|^2
+2s^2 \,\BB(T)  +  \tfrac{1}{3} s^4 \, \left\| T \right\|^4
\eee
where we used Lemmas \ref{ricTT} and \ref{TT4} in the last step. We conclude the proof
noting that 
\[
\sum_{i,j=1}^n \left(ric^A_{ij}\right)^2 =  s^2 \,
\sum_{i,j=1}^n \left(\delta T(E_i,E_j)\right)^2 = s^2\, \| \delta T \|^2.
\]
\hfill$\boxbox$
}

\begin{lemma}
Let $M$ be an $n$-dimensional Riemannian manifold.
For any $3$-form $T$ one has 
\[
\sum_{i,j,k=1}^n \ric^g_{ij} \, \left\langle T_{ik}, T_{jk} \right\rangle  = 
-  B(T) + \frac{1}{n} \, R^g \|T\|^2.
\]
\label{ricTT}
\end{lemma}
\pf{ We use the decomposition of the Ricci curvature into its trace-free
part and its trace (\ref{Ricci_decomposition}) and find
\bb
\sum_{i,j,k=1}^n \ric^g_{ij} \, \left\langle T_{ik}, T_{jk} \right\rangle
&=& \sum_{i,j,k=1}^n 
\B(E_i,E_j) \, \left\langle T_{ik}, T_{jk} \right\rangle
+ \frac{1}{n} \sum_{i,j,k=1}^n R^g \delta_{ij} \, \left\langle T_{ik}, T_{jk} \right\rangle
\nonumber \\
&=& - \sum_{i,j,k=1}^n 
\B(E_i,E_j) \, \left\langle T_{ik}, T_{kj} \right\rangle
+ \frac{1}{n} R^g \; \sum_{i,k=1}^n  \, \left\langle T_{ik}, T_{ik} \right\rangle
\nonumber \\
&=& -  B(T) + \frac{1}{n} \, R^g \|T\|^2
\eee
with $\BB(T)$ as defined in Theorem \ref{theoremriem}.
\hfill$\boxbox$
}
\begin{lemma}
Let $M$ be an $4$-dimensional Riemannian manifold.
For any $3$-form $T$ one has 
\[
\sum_{i,j,k,l=1}^4 \left\langle T_{ik}, T_{jk} \right\rangle
\left\langle T_{il}, T_{jl} \right\rangle = \tfrac{1}{3} \|T\|^4.
\]
\label{TT4}
\end{lemma}
\pf{
We compute:
\bb
\sum_{i,j,k,l=1}^4 \left\langle T_{ik}, T_{jk} \right\rangle
\left\langle T_{il}, T_{jl} \right\rangle 
&=& 
\sum_{i,j,k,l,n,m=1}^4 \ T(E_i,E_k,E_n) T(E_j,E_k,E_n)
T(E_i,E_l,E_m) T(E_j,E_l,E_m)
\nonumber \\
&=&  
\sum_{i,j,k,l,n,m=1}^4 \ T(E_n,E_k,E_i) T(E_m,E_l,E_i) 
T(E_n,E_k,E_j) T(E_m,E_l,E_j)
\nonumber \\
&=& 
\sum_{k,l,n,m=1}^4 \left\langle T_{kn}, T_{lm}\right\rangle^2.
\label{Tknlm}
\ee
We decompose $T=T_1 + T_2 + T_3 + T_4$ according to the definitions of the 
$T_a$ in Lemmas \ref{cct0} and \ref{Riemtttt}. 
Following precisely the arguments of Lemma \ref{Riemtttt} we obtain for (\ref{Tknlm}) from the Tables~\ref{tab1} and \ref{tab2} the summands given in Table~\ref{tab3}. It follows that
\bb
\sum_{k,l,n,m=1}^4 \left( \left\langle T_{kn}, T_{lm}\right\rangle \right)^2 = 12 \big( \sum_{a=1}^4 t_a^2  \big)^2.
\eee
Combining this with (\ref{T4}) (\ref{Tknlm})
finishes the proof.
\hfill$\boxbox$
}

\subsection{Proofs of Theorem \ref{Eulerallg} 
and Lemma \ref{intzero}}

\paragraph{Proof of Theorem \ref{Eulerallg}.}
Choose an orthonormal basis $E_1,..,E_4$ of $T_pM$ at a point $p$ and 
express the curvature $2$-form in this basis
\bb
\Riem_{ij} &=& \sum_{n,m=1}^4 \Riem_{ij}(E_n,E_m) (E_n)^\flat \wedge (E_m)^\flat
\nonumber \\
&=& \sum_{n,m=1}^4 \left\langle\Riem(E_i,E_j)E_n,E_m \right\rangle
(E_n)^\flat \wedge (E_m)^\flat
\nonumber \\
&=& \sum_{n,m=1}^4 \riem_{ijnm} (E_n)^\flat \wedge (E_m)^\flat
\eee
where $\riem$ is the $(4,0)$-tensor  defined in (\ref{Riem_decomp}). 
Now the $4$-form ${\bf K}$ reads
\bb
{\bf K} &=& \tfrac{1}{32 \pi^2} \sum_{i,..,p=1}^4 \epsilon_{ijkl} 
\riem_{ijnm} \riem_{klsp}
\, (E_n)^\flat \wedge (E_m)^\flat \wedge (E_s)^\flat \wedge (E_p)^\flat
\nonumber \\
&=& \tfrac{1}{32 \pi^2} \sum_{i,..,p=1}^4 \epsilon_{ijkl} \epsilon_{nmsp}
\riem_{ijnm} \riem_{klsp} \, dvol.
\eee
With the standard equality
\[
\epsilon_{ijkl} \epsilon_{nmsp} = \det 
\begin{pmatrix}
\delta_{in} & \delta_{im} & \delta_{is} & \delta_{ip} \\
\delta_{jn} & \delta_{jm} & \delta_{js} & \delta_{jp} \\
\delta_{kn} & \delta_{km} & \delta_{ks} & \delta_{kp} \\
\delta_{ln} & \delta_{lm} & \delta_{ls} & \delta_{lp} 
\end{pmatrix}
\]
and $\ric_{ij} = \ric(E_i,E_j)$ one finds 
\bb
{\bf K} = \tfrac{1}{8 \pi^2} \left( R^2 - 4 \, \sum_{j,k=1}^4 \ric_{ij} \, 
\ric_{ji} \, + \sum_{i,j,k,l=1}^4 \riem_{ijkl} \, \riem_{klij}
\right) \, dvol.
\eee
Decomposing the Riemann and the Ricci curvature into their 
symmetric and anti-symmtric components we find
\bb
\sum_{j,k=1}^4 \ric_{ij} \, \ric_{ji} &=&
\sum_{j,k=1}^4 (\ric^S_{ij})^2  - \sum_{j,k=1}^4 (\ric^A_{ij})^2 
- \sum_{j,k=1}^4 \ric^S_{ij} \, \ric^A_{ij}  
+ \sum_{j,k=1}^4 \ric^A_{ij} \, \ric^S_{ij}  
\nonumber  \\
&=& \| \ric^S \|^2 - \|\ric^A \|^2
\eee
and by the same argument
\[
\sum_{i,j,k,l=1}^4 \riem_{ijkl} \, \riem_{klij}
= \| \riem^S \|^2 - \|\riem^A \|^2
\] 
Thus we have 
\[
{\bf K} = \tfrac{1}{8 \pi^2} \left( R^2 - 4 \| \ric^S \|^2 + 4 \|\ric^A \|^2 +
\| \riem^S \|^2 - \|\riem^A \|^2 \right) \, dvol.
\]

\hfill$\boxbox$
\medskip

\begin{cor}
Let $M$ be a closed $4$-dimensional manifold with Riemannian metric $g$,  $\nabla$
an arbitrary orthogonal connection and  $\nabla^g$ the Levi-Civita connection. Then
\bb
\int_M \left((R^g)^2 - 4\, \| \ric^g \|^2 + \| \riem^g \|^2 \right)\, dvol
&=&
\int_M \left(R^2 - 4\, \| \ric^S \|^2 
+ 4\, \| \ric^A \|^2 \right.
\nonumber \\
&& \qquad + \left. \| \riem^S \|^2 -  \| \riem^A \|^2 \right)\, dvol  
\eee
\label{corvanish}
\end{cor}
\pf{
The Euler characteristic is a topological invariant and does not depend on the 
choice of the connection in the representation of the Euler class.
\hfill$\boxbox$
}
\medskip

\noindent

\paragraph{Proof of Lemma \ref{intzero}.}
Let  $\riem$, $\ric$ and $R$ be 
the Riemann curvature, the Ricci curvature and the scalar curvature 
of $\nabla$. With the decomposition of the norm squared of the symmetric
part of the 
Riemann curvature, theorem \ref{theoremriem}, the 
the Ricci curvature, theorem \ref{theoremric}, and   
the scalar curvature,equation (\ref{Scalar_identity}), 
we have from corollary \ref{corvanish} 
\bb
0 &=& \int_M \left(R^2 - (R^g)^2  - 4\,( \| \ric^S \|^2 - \| \ric^A \|^2 - \, \| \ric^g \|^2)
+ \| \riem^S \|^2 -\| \riem^A \|^2 - \| \riem^g \|^2 \right)\, dvol 
\nonumber \\
&=&   \int_M \left( - 2 s^2 R^g \, \|T \|^2 + s^4 \| T \|^4  
-  \tfrac{4}{3} s^4 \left\| T \right\|^4  
+ 2 s^2 \,R^g\,\left\|T\right\|^2
- 8 s^2  \,\BB(T)  + 4 s^2 \| \delta T \|^2 \right)\, dvol
\nonumber \\
&& + \int_M  \left( \tfrac{1}{3} s^4 \,  \left\| T \right\|^4 
+ \tfrac{1}{4} s^2 \, \left\|dT \right\|^2 
- \tfrac{1}{3} s^2 \,R^g\,\left\|T\right\|^2
+4 s^2 \,\BB(T)  - \left\|\riem^A \right\|^2  \right)\, dvol
\nonumber \\
&=& s^2  \int_M \left( 4  \| \delta T \|^2  - \tfrac{1}{3}\,R^g\,\left\|T\right\|^2  
+ \tfrac{1}{4} \left\|dT \right\|^2 - 4\,\BB(T) - \frac{1}{s^2} \left\|\riem^A \right\|^2 \right)\, dvol.
\eee
For any $s\neq 0$ follows the assertion of the lemma. 
\hfill$\boxbox$

\begin{rem}
At first glance the cancelation of the term $\|T\|^4$ in the proof of Lemma \ref{intzero} might surprise.
But it has to vanish because it is the only term scaling in the $4^{th}$ power in the  pre-factor $s$ 
while all other terms scale quadradically in $s$.
\end{rem}
%

%

\end{document}